\documentclass[aps,prc,preprint,groupedaddress,superscriptaddress]{revtex4}
\usepackage[dvipdfmx]{graphicx}
\usepackage{ulem}

\begin{document}


\title{Self-consistent calculation of  the
nuclear composition in hot and dense stellar matter}


\author{Shun Furusawa}
\email{furusawa@fias.uni-frankfurt.de}
\affiliation{Frankfurt Institute for Advanced Studies, J.W. Goethe University, 60438 Frankfurt am Main, Germany}
\author{Igor Mishustin}
\affiliation{Frankfurt Institute for Advanced Studies, J.W. Goethe University, 60438 Frankfurt am Main, Germany}
\affiliation{Russian  Research Center Kurchatov Institute, Moscow 123182, Russia
}


\date{\today}

\begin{abstract}
We investigate the mass fractions  and in-medium properties 
 of  heavy nuclei in stellar matter at characteristic
 densities and temperatures  for supernova (SN) explosions.  
The individual nuclei are described within the compressible liquid-drop model taking into account 
modifications of bulk, surface and Coulomb energies. 
The equilibrium properties of nuclei   
and the full ensemble of heavy  nuclei  are calculated self-consistently.   
It is found that heavy nuclei in the  ensemble are either compressed or decompressed 
depending on the isospin asymmetry of the system.
The compression or decompression  has a little influence on the binding energies, total mass fractions and average mass numbers of heavy nuclei, although the equilibrium densities of individual nuclei  themselves are changed appreciably above one hundredth  of normal nuclear density.
We find that  nuclear structure in single nucleus approximation deviates from the actual one obtained in the multi-nucleus description,
since the density of free nucleons is different between these two descriptions. 
This study indicates that 
a  multi-nucleus description is required to realistically account for 
in-medium effects on  the nuclear structure  in supernova matter. 
\end{abstract}


\maketitle

\section{Introduction \label{intro}}
Hot and dense matter can be realized in  core collapse supernovae and mergers of compact stars.
The details of these phenomena are not completely clarified yet
because of their complexity 
\cite{janka12,kotake12,burrows13,foglizzo15,shibata11,faber12}.
One of the central problems 
 is the nuclear equations of state (EOSs) of the hot and dense matter both at sub- and supra-nuclear densities. 
The EOS provides information on composition of nuclear matter in addition to thermodynamical quantities such as pressure and entropy.
The nuclear composition plays an important role to determine the evolution of the lepton fraction through weak interactions \cite{raduta16,furusawa17b}. 
This lepton fraction is  one of the most critical ingredients for the dynamics and  synthesis of heavy elements
in  these events  \cite{hix03,lentz12,wanajo14, sekiguchi15}.

There are two types of EOSs for the  supernova and merger simulations.
The single nucleus approximation (SNA) is the option, in which
 the ensemble of heavy nuclei is represented by a single nucleus.
Two  standard EOSs widely used for the simulations of core-collapse  supernovae,
Lattimer and Swefty \cite{lattimer91} and Shen et al.   \cite{shen98a,shen98b,shen11},
belong to the category.
They calculated  the  representative heavy nucleus described by the compressible liquid drop model   and Thomas Fermi model,  respectively. 
In such calculations, 
 in-medium effects on a single nucleus, such as compression and deformation, are taken into account more easily 
in comparison with the multi-nucleus EOS.
On the other hand, they are not able to  provide a nuclear composition, which is needed to estimate weak interaction rates in supernova and merger simulations.
Furthermore,  even the average mass number and total mass fraction of heavy nuclei
may not be correctly reproduced by the representative nucleus alone \cite{burrows84,hempel10,furusawa11}.

The other option is multi-nucleus approximation (MNA), in which the full ensemble of nuclei  is obtained for each set of thermodynamical condition.
In recent years, some EOSs with MNA have been formulated by different research groups.
SMSM EOS \cite{botvina04, botvina10,buyukcizmeci14} is  a generalization of the Statistical 
 Multi-fragmentation Model (SMM)
successfully used for the description of nuclear fragment produced in
heavy-ion collisions \cite{bondorf95}.
In this model, temperature  dependences of bulk and surface energies are taken into account using the incompressible liquid-drop model (LDM).
Hempel et al. have also constructed an  EOS (HS EOS) 
based on the relativistic mean field theory (RMF) \cite{hempel10}.
In this model, nuclear binding energies as well as nuclear shell effects are based on experimental and theoretical mass data.
In the FYSS EOS  \cite{furusawa11,furusawa13a, furusawa17a},
 the mass formula extended from the 
 LDM  was used to describe  nuclear shell effects  as well as various in-medium effects.
As demonstrated in Buyukcizmeci  et al. \cite{buyukcizmeci13} by direct comparison, 
there are very large differences in nuclear composition predicted by these three models, especially at high densities.
A hybrid EOS has been provided by G. Shen et al. \cite{sheng11}, 
where a multi-nucleus EOS based on the virial expansion at low densities is matched to 
a single-nucleus EOS via  Hartree approximation at high densities, i.e., 
MNA  is employed only in the low density regime.
An interesting approach has been developed recently \cite{typelpc},
where the Thomas-Fermi description of individual nuclei was confined with the full nuclear ensemble calculations.

The purpose of the present study is to investigate  in-medium effects in multi-nucleus  description.
The in-medium effects on nuclear binding energies are investigated mainly in SNA;
Papakonstantinou et al. \cite{papakonstantinou13}  have used the  local density approximation
and Aymard \cite{aymard14}  
 Extended Thomas-Fermi Approximation.
They showed that the binding energies strongly depend on the density and asymmetry of the nucleons in vapor. 
On the other hand, multi-nucleus EOSs
use very simplified  approximations
for nuclear biding energies  at high densities and temperatures  to calculate the statistical  ensemble.
The fully self-consistent calculation with the multi-nucleus composition 
including in-medium effects on  nuclear structure has not been done yet due to the complexity.
Gulminelli and Oertel \cite{gulminelli15} have investigated 
the equilibrium between nucleons and nuclei as well as nuclear component in multi-nucleus descriptions, ignoring
the in-medium  modification of nuclear structure due to the 
compression and surface interactions with surrounding nucleons.

In the present paper, we take into account self-consistently the compression or decompression of heavy nuclei in the multi-nucleus description, i.e., the changes of equilibrium densities of individual nuclei embedded in dense stellar environment. Then we evaluate the corresponding changes in the nuclear abundances by the optimization of these equilibrium densities. The comparison of the nuclear equilibration in single- and multi-nucleus descriptions is also done.

We consider the present work as an attempt to discuss and evaluate several new effects which may appear in dense stellar matter and which have not being analyzed previously.
The new EOS we construct in this study  is not aimed at the immediate practical use
 in astrophysical simulations,
since the required self-consistent calculations need huge computational resource. 
To reduce the computational cost,  the nuclear interactions among nucleon vapor outside nuclei are not considered in the present study,
although they are included  in many SN EOS via RMF or  Skyrme type interactions. 
In addition, we ignore the shell effect and light clusters other than the alpha particles. 
The former  is  very important to estimate the weak rates at low temperatures  $T \lesssim 2$ MeV \cite{furusawa17b}
 and the latter also may have influence on the supernova dynamics  \cite{furusawa13b,fischer16}.
 On the other hand, their theoretical consideration is rather naive and  many uncertainties still remain
\cite{buyukcizmeci13,furusawa13a,hempel11,oertel16}.
In fact at present, nobody can calculate the exact EOS for hot and dense stellar matter starting from first principles.
Therefore, we are using a simplified approach which should be considered as a preliminary step for the 
future work to construct practical and more realistic EOSs for astrophysical simulations.

This article is organized as follows.
In section~\ref{sec:model}, we describe our model of EOS,
which attempts to solve some of the problems mentioned above.
The results  are shown in section~\ref{sec:res}.
The paper is wrapped up with a conclusion in section~\ref{sec:conc}.

\section{Model \label{sec:model}}
Total free energy density 
 is represented as:
\begin{eqnarray}
f = f_p + f_n  + f_{\alpha}+ \sum_i  n_i (F^t_{i} + M_{i}) ,
\label{total}
\end{eqnarray}
where $f_p$, $f_n$  and $f_{\alpha}$ are the free energy densities of free protons, neutrons and alpha particles  and $n_{i}$,
$F^t_{i}$ and $M_{i}$ are the number densities, 
translational free energies and rest masses of individual heavy nuclei $i$ with the proton number $6 \leq Z_i \leq 1000$.
In this study, the light nuclei ($Z \leq 5$) are represented by $\alpha$-particles.
 Other light clusters are ignored, as is usually done in single nucleus models. 
This is also needed to make  more fair  the comparison between  MNA and SNA descriptions.
The  EOS as a function of baryon density $n_B$,  temperature $T$ and total proton fraction $Y_p$  
is obtained by minimizing the free energy density.

Nucleons are considered as structure-less point-like particles moving in the excluded-volume: $V-V_N$, 
 where $V$ is  the total volume,  $V_N$ is the volume occupied by heavy nuclei, $V_N=\sum_i A_i/n_{si}$ and  $n_{si}$ and  $A_{i}$  are the equilibrium density and the mass number of nucleus $i$.
The local number densities of free protons and neutrons are defined as $n'_{p/n}=(N_{p/n})/(V-V_N)$  with the numbers of free protons, $N_p$, and  free neutrons, $N_n$.
The free energy density  of free nucleons is obtained  as 
\begin{equation}%
\label{eq:pn}
f_{p/n}=(1-V_N/V) n'_{p/n}  T \left\{ \log \left(\frac{n'_{p/n}}{g_{p/n}   (m_{p/n} T/2\pi \hbar ^2  )^{3/2} }  \right)- 1 \right\} \  
\end{equation}
with 
 $g_{p/n}=2$ and  
the nucleon masses  $m_{p/n}$.
This ideal gas approximation fails at high densities and low temperatures  due to nuclear interactions and quantum statistics. However, we believe that the results will not be qualitatively changed, as discussed later.
The translational free energy of nuclei $i$ in Eq.~(\ref{total}) is obtained using the expression for the ideal Boltzmann gas with excluded volume correction
\begin{equation}%
\label{eq:tra}
 F_{i}^{t} = T \left\{ \log \left(\frac{ n_{i}/\kappa}{g_{i} (M_{i} T/2\pi \hbar ^2 )^{3/2}  }\right)- 1 \right\} , 
\end{equation}
where  
$g_i$ is the spin degeneracy factors of the ground state, which is set to one,
and $\kappa=1-n_B/n_0$. 

The calculation  below is based  on the mass formula used in SMSM and FYSS EOSs,
where  an additional modification in the symmetry energy has been introduced
(parameter $L$,  see below).
In this study for simplicity, we neglect
the nuclear shell  effects, which are derived from
nuclear structure calculations of ground states but should be washed out  at $T\sim 2-3$~ MeV \cite{furusawa17a}.
The internal free energy of  heavy nuclei is assumed to be the sum of  bulk, Coulomb and surface free energies: 
\begin{equation}%
F_i=F_i^B+F_i^C+F_i^{S} .
\end{equation}%
We find the equilibrium density of each  nucleus, $n_{si}$, 
   at given $T,Y_p,n'_p,n'_n$
as the baryon density, at which the internal free energy of  the nucleus takes its minimum value. The nuclear mass in Eq.~(\ref{total}) is defined as 
\begin{equation}%
M_i=Z_i m_p + (A_i -Z_i)  m_n + F_i ,
\end{equation}%
where $F_i$ is taken at the equilibrium density.

The bulk free energy of nucleus $i$ is evaluated as
\begin{equation} \label{eq:bulk}
F_{i}^{B}(n_{si},T) = A_i \omega(n_{si},Z_i/A_i,T)    , 
\end{equation}
where $ \omega(n_{B},x,T)$ is the free energy per baryon of bulk nuclear matter as a function  of baryon density in nuclei $n_B$, 
charge fraction  $x=Z/A$ and temperature $T$, which is expressed as 
\begin{equation}%
\label{eq:omega}
 \omega(n_B,x,T)= \omega_0 + \frac{K_0}{18 n_0^2} (n_B-n_0)^{2}   + \left[S_0 +\frac{L}{3 n_0}  (n_B-n_0)   \right] (1-2x)^2  - \frac{T^2}{\epsilon_0}  , 
 \label{eq:para}
\end{equation}
with 
 $\epsilon_0=16$~MeV
\cite{lattimer91,bondorf95}.
We adopt the parameters for bulk properties at zero temperature, 
$\omega_0$, $n_0$, $K_0$, $S_0$ and $L$ from Refs. \cite{oyamatsu03,oyamatsu07},
 which are summarized in Tab.~\ref{tab1_bulk}. 
The parameter set E is the most standard parameter set that reproduces 
the nuclear masses and radii of stable nuclei in the Thomas Fermi Calculations \cite{oyamatsu03}.
 They are used in most calculations in this study, while the comparisons between the parameter sets B and E are also presented. 
The thermal term, $T^2/\epsilon_0$, comes from the integration over excited states of nuclei,
which can be taken in low-temperature approximation at  $T \ll \epsilon_F$, where $\epsilon_F$ is the Fermi energy $\sim$ 40 MeV at saturation density.
Here the slope parameter $L$ 
accounts for the density dependence of the symmetry energy of nuclei
when their equilibrium density deviates from $n_0$.
 Its actual value
 is still rather uncertain. 
Various predictions have been provided by the terrestrial nuclear experiments, theoretical analyses and astrophysical observations  \cite{tsang12,newton13,lattimer14a,hebeler14,danielewicz14,sotani15}.
Unfortunately, these constraints do not overlap and 
the actual value is likely to be larger than 35 MeV  and less than 131 MeV.
The equilibrium densities of asymmetric nuclei are sensitive to $L$: the larger $L$ leads to  the smaller saturation density of asymmetric bulk matter,
\begin{equation}
\label{eq:nsbulk} 
n_{s,bulk}(x)=n_0-3n_0L(1-2x)^2/K_0,
\end {equation}
at which $\omega$ takes its minimum value.
The other parameters have also some uncertainties as
$n_0=0.15-0.16$ fm$^{-3}$,
 $K_0= 240\pm20$ MeV 
and $S_0=29.0-33.7$ MeV \cite{shlomo06,lattimer13,danielewicz14,oertel16}.

The Wigner-Seitz (WS)  cell  for each species of  nuclei $i$  is set to satisfy the charge neutrality 
 within the volume, $V_i$.
The cell also contains free nucleons as a vapor outside the nucleus as well as  electrons distributed  uniformly in the entire cell.
The charge neutrality in the cell gives the cell volume
\begin{equation}
V_i = (Z_i - n'_p  V_{Ni})/(n_e-n'_p) ,
\end{equation}
where $V_{Ni}$ is the volume  of the nucleus in the cell and can be calculated as $V_{Ni} = A_i / n_{si}$ and 
 $n_e=Y_p n_B$  is  the number density of electrons. 
The expression for the Coulomb free energy  is 
obtained within the WS approximation
 by integrating  the Coulomb potential over the cell $i$  containing nucleus $(A_i,Z_i)$
\begin{eqnarray}
F_i^C(n_{si},n'_p,n_e)= \frac{3}{5} \left(\frac{3}{4 \pi} \right)^{-1/3}  e^2 n_{si}^2 \left(\frac{Z_i - n'_p V_{Ni}}{A_i}\right)^2  {V_{Ni}}^{5/3} D(u_i), 
\label{eqclen}
\end{eqnarray} 
where $u_i=V_{Ni}/V_i$, $D(u_i)=1-\frac{3}{2}u_i^{1/3}+\frac{1}{2}u_i$ and $e$ is the elementary charge. 
Note that the formation of nuclear pasta phases is not taken into consideration in this study. They are essential as intermediate state at the transition to uniform nuclear matter \cite{watanabe05,newton09,okamoto12,schneider13,horowitz16}. Although their realistic calculation in MNA is rather complicated, they are taken into account phenomenologically in the FYSS EOS. Also, the electron distribution deviate from the uniform one, especially in the nuclear pasta phases  \cite{maruyama05,ebel15}. However, in this paper, we address the densities at $n_B \lesssim 0.3 n_0$, where these effects are not very important.

The surface free energy is evaluated 
by the expression generating the one form  \cite{bondorf95},
\begin{eqnarray}
F_i^{S} (n_{si},n'_p,n'_n,T) = 4 \pi {r^2_{Ni}} \, \sigma_i \left(1-\displaystyle{\frac{n'_p+n'_n}{n_{si}}} \right)^2 \left( \displaystyle{\frac{T^2_c-T^2}{T^2_c+T^2}} \right)^{5/4}  ,   \label{eq:surf} 
\end{eqnarray} 
where  $r_{Ni} = ( 3/4 \pi V^N_i)^{1/3}$  is the radius of nucleus, 
$T_c=18$ MeV is  the critical  point,  at which the distinction  between  the  liquid  and  gaseous  phase  disappears, and
\begin{eqnarray}
\sigma_{i} =\sigma_0 \displaystyle{\frac{16 + C_s}{ (1-Z_i/A_i)^{-3} + (Z_i/A_i)^{-3}  +C_s} } ,  \label{eq:surf2}
\end{eqnarray} 
where $\sigma_0$ denotes the surface tension for symmetric nuclei. 
 The values of the constants, $\sigma_0=1.14$ MeV/fm$^3$ and 
$C_s =$ 12.1 MeV, are
taken  from  Agrawal et al. \cite{agrawal14b}. 
The last two factors  in Eq.~(\ref{eq:surf})
describe  the reduction of the surface free energy  
 when the density contrast between the nucleus and the nucleon vapor  decreases \cite{furusawa11}, and when the temperature grows.

 The free energy density of alpha particles is represented as $f_{\alpha}=n_{\alpha} (F_{\alpha}^{t} +M_{\alpha})$, where the translational energy is calculated by the same formula as for heavy nuclei, Eq.~(\ref{eq:tra}),  but the mass is set to be the experimental value without any modifications at high temperature and densities. 

The abundances of nuclei as a function of $n_B$, $T$ and $Y_p$ are obtained by minimizing the model free energy with 
respect to the number densities of nuclei and nucleons under the constraints,
\begin{eqnarray}
 n_p+n_n+ 4 n_{\alpha}+\sum_i{A_i n_i} & = & n_B, \label{eq:cons1} \\
 n_p+ 2 n_{\alpha}+\sum_i{Z_i n_i} & = & n_e=Y_p n_B . \label{eq:cons2}
\end{eqnarray}
In our EOS, the free energy density of nuclei depends on the local number densities of protons and neutrons $n'_{p/n}$. 
The nucleon chemical potentials  $\mu_{p/n}$  are expressed as follows:
\begin{equation}
\mu_{p/n} =\frac{\partial f}{ \partial n_{p/n}}= \mu'_{p/n}(n'_p,n'_n) + \sum_{i} n_{i} {\frac{\partial F_{i}(n'_p,n'_n)}{ \partial n_{p/n}}} ,   
\label{eq:chem}
\end{equation}
where $n_{p/n}=N_{p/n}/V$,   $\mu'_{p/n}={\partial f_{p/n}}/{\partial n'_{p/n}}$ are the chemical potentials of nucleons in the vapor and the second term originates from the dependence of the free energies of nuclei on $n'_{p/n}$. 
We hence solve  the equations relating $\mu_{p/n}$ and $n'_{p/n}$,
 Eq.~(\ref{eq:chem}), 
 as well as the two constraint equations, Eqs.~(\ref{eq:cons1}) and (\ref{eq:cons2}),
to determine the four variables: $\mu_p$, $\mu_n$, $n'_p$ and $n'_n$.

For comparison,  in addition to the compressible liquid drop model
 in multi-nucleus description (Model MC), 
which has been explained above in this section,
 we  have considered two other simplified EOSs.
The first EOS corresponds to the incompressible liquid drop model in multi-nucleus description  (Model MI),
where the  equilibrium density  of each nucleus is assumed to be fixed to the corresponding value  in vacuum, 
 i.e. $n_{si}(T,n_e,n'_p,n'_n)=n_{si}(0,0,0,0)$.  
The other is single nucleus EOS with compressible liquid drop model  (Model SC), in which 
the  mass and proton numbers of a representative nucleus, $A_{rep}$ and $Z_{rep}$, are introduced  instead of an ensemble of nuclei.
In this case, the free energy density is expressed as  $f = f_{p} +f_{n} + f_{\alpha} + n_{rep} (F^t_{rep} + M_{rep})$ instead of Eq.~(\ref{total}),
where $n_{rep}$, $F^{t}_{rep}$ and $M_{rep}$ are the number density, translational free energy and mass  of representative nucleus. 
To find 6 unknowns, $\mu_{p}$, $\mu_{n}$, $n'_{p}$, $n'_{n}$, $A_{rep}$ and $Z_{rep}$,
we use the  constraints of Eqs.(\ref{eq:cons1}-\ref{eq:chem})
and two definitions ${\partial f}/{ \partial A_{rep}}|_{Z_{rep}}=n_{rep} \mu_n$ and ${\partial f}/{ \partial Z_{rep}}|_{A_{rep}}=n_{rep} (\mu_{p}-\mu_{n})$.
The former and latter may be regarded as surrogates for
 the multi-nucleus EOSs (SMSM, HS, FYSS  EOSs) and  single nucleus EOSs (STOS, LS EOSs) for supernova matter,
 although various parts in  free energies of nucleons  and nuclei used by different authors are quite different from each other.

\section{Results  \label{sec:res}}
\begin{figure}
\includegraphics[width=8.1cm]{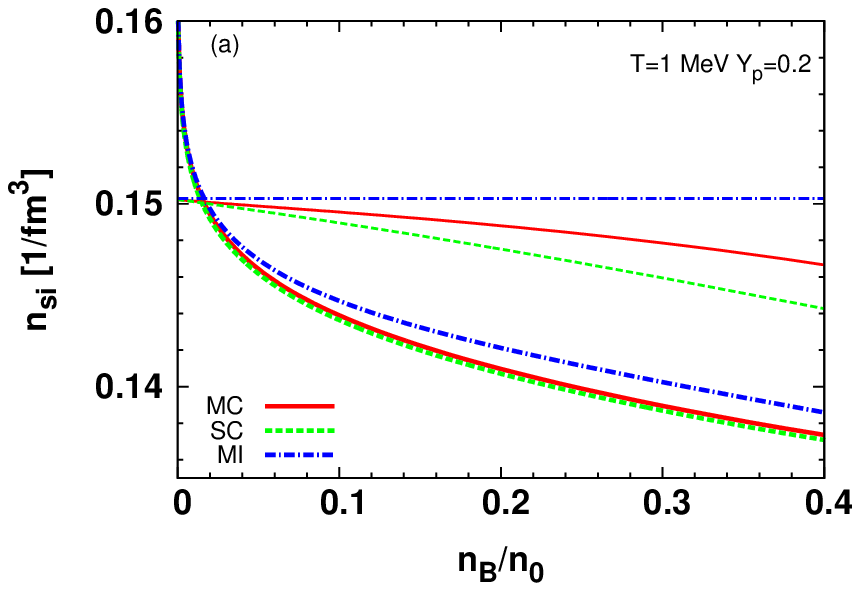}
\includegraphics[width=8.1cm]{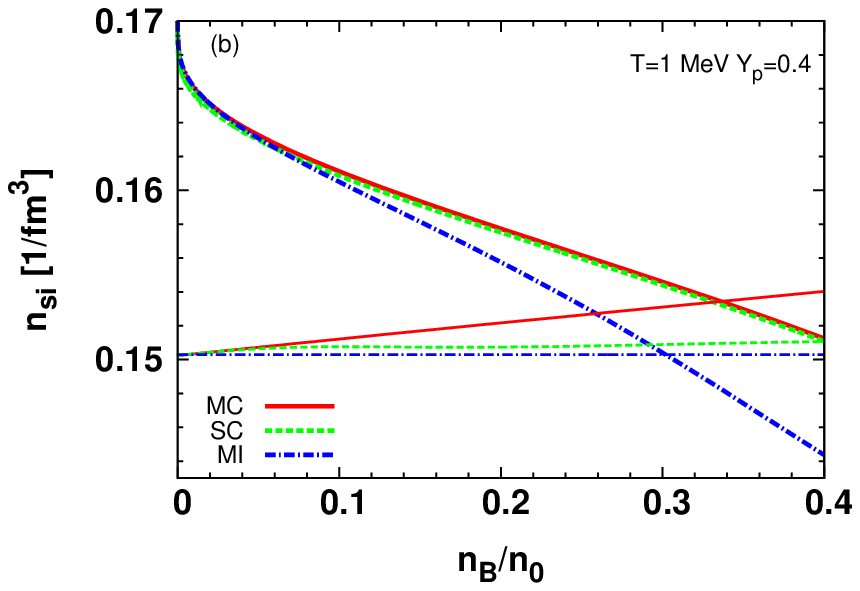}
\includegraphics[width=8.1cm]{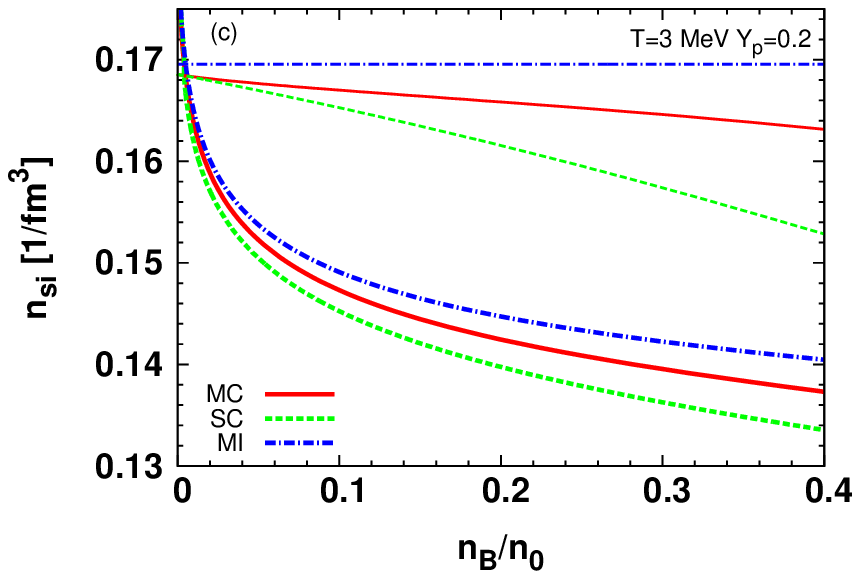}
\includegraphics[width=8.1cm]{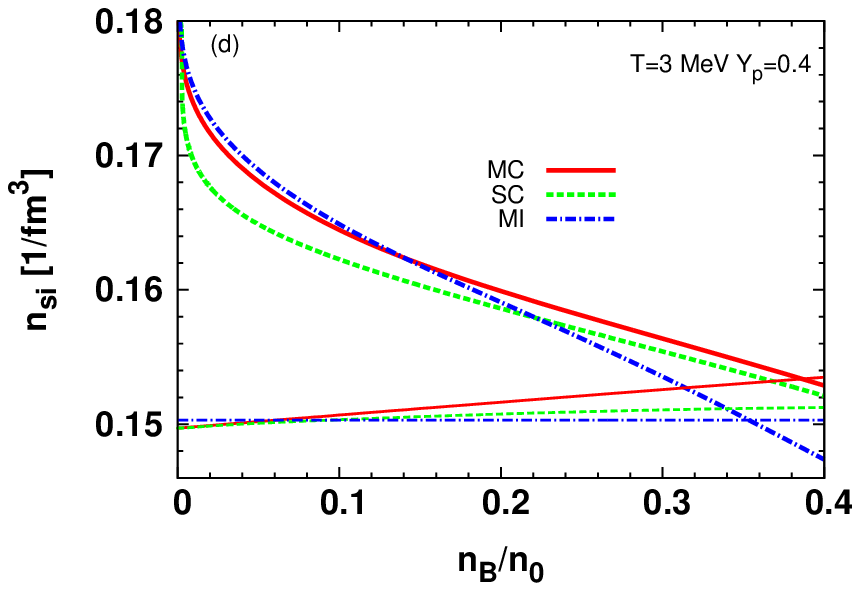}
\caption{Average  equilibrium density of heavy nuclei 
for
the compressible liquid drop model
(Model MC, red solid thick lines) and  incompressible one (Model MI, blue dashed-dotted thick lines)  in the multi-nucleus descriptions
 and  the  equilibrium density of representative nucleus for  
the single nucleus EOS with compressible liquid drop model
 (Model SC, green dashed  lines)
 as a function of  $n_B$
at  $T=1$ MeV (top row) and 3 MeV (bottom row)
and $Y_p=$ 0.2 (left column) and 0.4 (right column). 
Thin lines indicate  individual  equilibrium densities of
$^{50}_{20}$Ca at $(T, Y_p)$=~(3 MeV, 0.2)
and  of  $^{300}_{100}$Fm at the other conditions
for
 Models MC  (red solid lines) and MI (blue dashed-dotted lines). 
 Green dashed thin lines show results for the
single nucleus approximation, 
in which the representative nucleus is
 fixed as $^{50}$Ca or  $^{300}$Fm 
   at any density. 
}
\label{fig_satuden}
\end{figure}

Below we present reults for the typical stellar conditions
 $(T, Y_p)$=~(1 MeV, 0.2), (1 MeV, 0.4), (3 MeV, 0.2), and (3 MeV, 0.4).
Figure~\ref{fig_satuden} displays the equilibrium density of representative 
nucleus in single nucleus EOS and  the average values for multi-nucleus ensembles.
On the whole, the  ensemble-averaged equilibrium densities decrease as the baryon density increases,
since the nuclei with larger mass numbers  are allowed, which have smaller equilibrium densities.
The  equilibrium densities of 
the nuclei with $(A_i,Z_i)=(50,20)$, $^{50}_{20}$Ca, for $(T, Y_p)$=~(3 MeV, 0.2)
and with $(A_i,Z_i)=(300, 100)$,  $^{300}_{100}$Fm, for the other conditions
 are also shown in Fig.~\ref{fig_satuden}. 
In the SNA,  
the representative nucleus at any density is assumed to be the nucleus with $(A_{rep},Z_{rep})=$
(50,20) or (300,100) described by the compressible liquid drop model.
In the MNA, $n_{si}$ of $^{50}$Ca or   $^{300}$Fm
is obtained by calculating the whole ensemble of nuclei. 
We find  
that the individual nucleus in  Model MC is compressed 
for  $Y_p$=~0.4 due to the reduction of  Coulomb free energy by the surrounding electrons
as the baryon density increases.
On the other hand, in a neutron-rich system with $Y_p$=~0.2, the equilibrium
 densities of  nuclei  decrease due to the two reasons: a) 
 the impact of electrons on Coulomb  free energies is less strong than for  $Y_p=0.4$ and  b) surface energies of nuclei
are reduced because of  the dripped neutrons.
 At non-zero temperatures, surface free energies of nuclei are reduced by the factor $((T^2_c-T^2)/(T^2_c+T^2))^{5/4}$ in Eq.~(\ref{eq:surf2}) even at zero density, leading to  the  equilibrium densities smaller than those in vacuum.
 Hence, the equilibrium densities in Model MC are smaller than those in Model MI at $T=3$~MeV  and $n_B=0$. 
It is found that the equilibrium densities in Model SC both for representative nucleus and for  $^{50}$Ca or   $^{300}$Fm  are smaller  than those in multi-nucleus EOSs in most cases, since more free nucleons are produced outside the nuclei, reducing the surface energies.

\begin{figure}
\includegraphics[width=12cm]{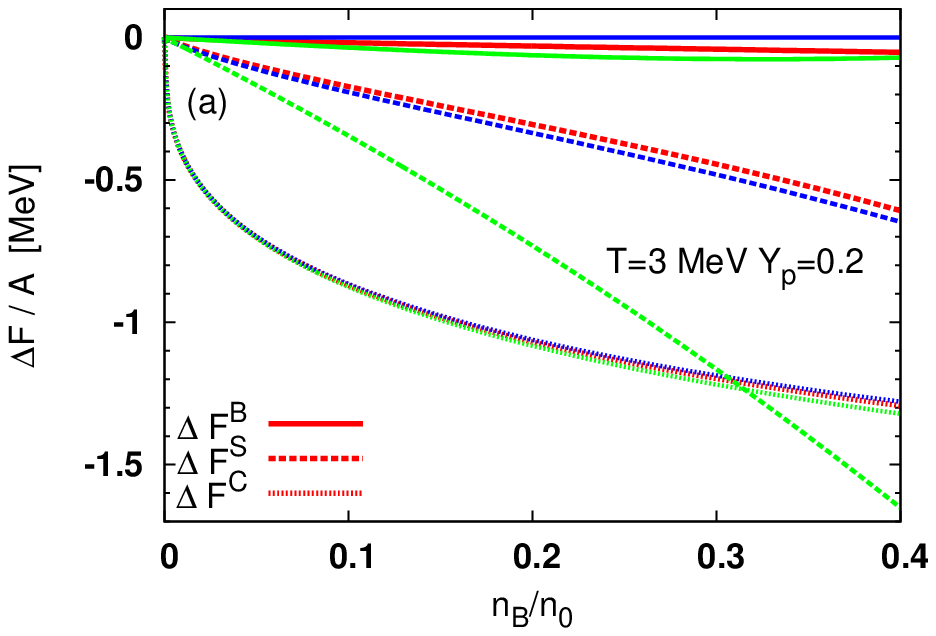}
\includegraphics[width=12cm]{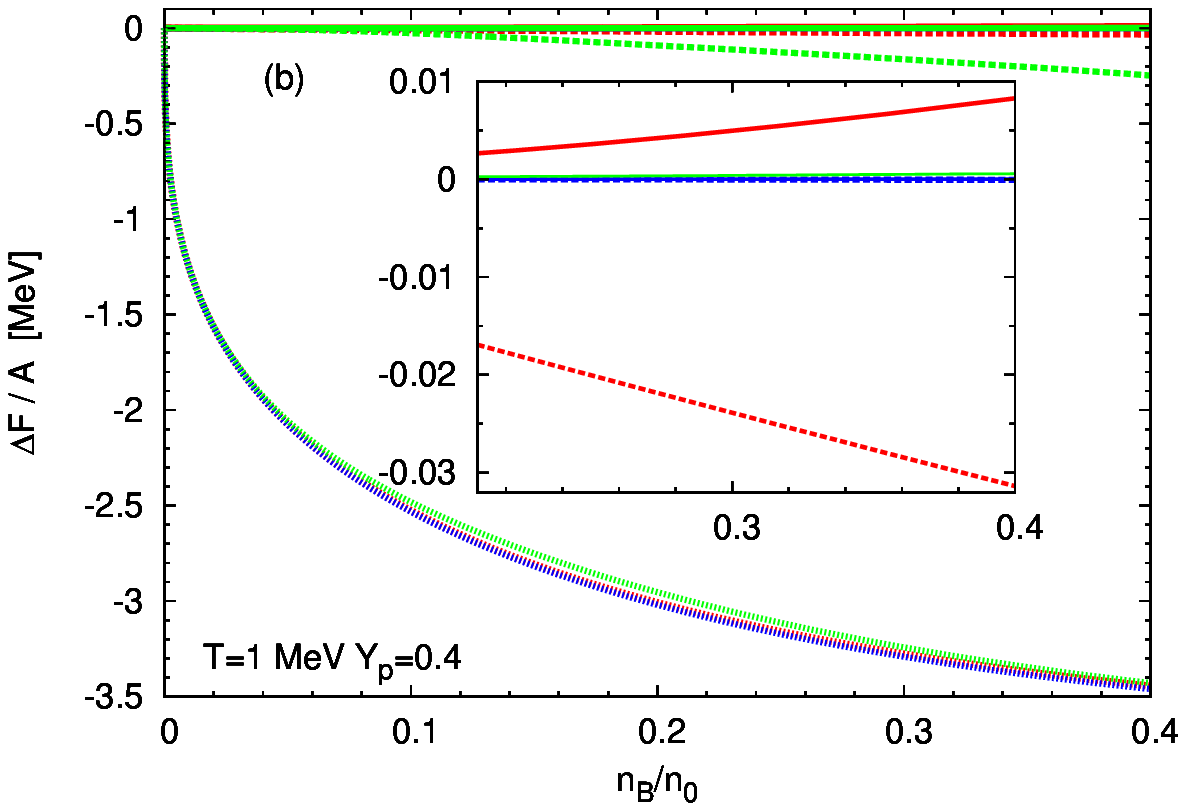}
\caption{Energy shifts per baryon  in bulk (solid lines), surface  (dashed lines) and Coulomb (dotted lines) terms, $ \{ F^{B/S/C}_{i}(T,n_e,n'_p,n'_n)-F^{B/S/C}_{i}(T,0,0,0)\}/A_i$,  of  $^{50}$Ca  at $(T, Y_p)$=~(3 MeV, 0.2)  (top)
and of $^{300}$Fm  at $(T, Y_p)$=~(1 MeV, 0.4)  (bottom)
  for Models MC  (red lines), MI (blue lines)  and SC (green lines).
}
\label{fig_bindshift}
\end{figure}

The binding energies of nuclei as well as  their equilibrium densities are essentially affected by electrons and free nucleons, especially at high baryon densities. 
Fig.~\ref{fig_bindshift} displays 
the shifts of bulk, surface and Coulomb free energies  of 
$^{50}$Ca 
 for $(T, Y_p)$=~(3 MeV, 0.2)
and 
of  $^{300}$Fm 
 for $(T, Y_p)$=~(1 MeV, 0.4),
 $\Delta F^{B/S/C}_{i}= F^{B/S/C}_{i}(T,n_e,n'_p,n'_n)-F^{B/S/C}_{i}(T,0,0,0)$. 
The bulk free energies are sensitive to the equilibrium density.
Those in Model MI are constant, whereas those in Model MC decrease 
for $(T, Y_p)$=~(3 MeV, 0.2)  and  increase  for $(T, Y_p)$=~(1 MeV, 0.4) due to  the decompression and  compression, respectively.
The bulk free energy of representative nucleus in Model SC is  always smaller than that in Model MC,
 since its equilibrium density is smaller. 
Note that  
the representative nucleus  is assumed to be
$^{50}$Ca or   $^{300}$Fm at any density here.
In all models, the  surface and Coulomb free energies are reduced by free nucleons and  electrons, respectively.
The equilibrium densities  also affect  them, as can be seen from the comparison of Models MC and MI, namely
 the surface free energies are increased and   Coulomb free energies are decreased  by decompression for  $(T, Y_p)$=~(3 MeV, 0.2)  and  vice versa for $(T, Y_p)$=~(1 MeV, 0.4).

\begin{figure}
\includegraphics[width=12cm]{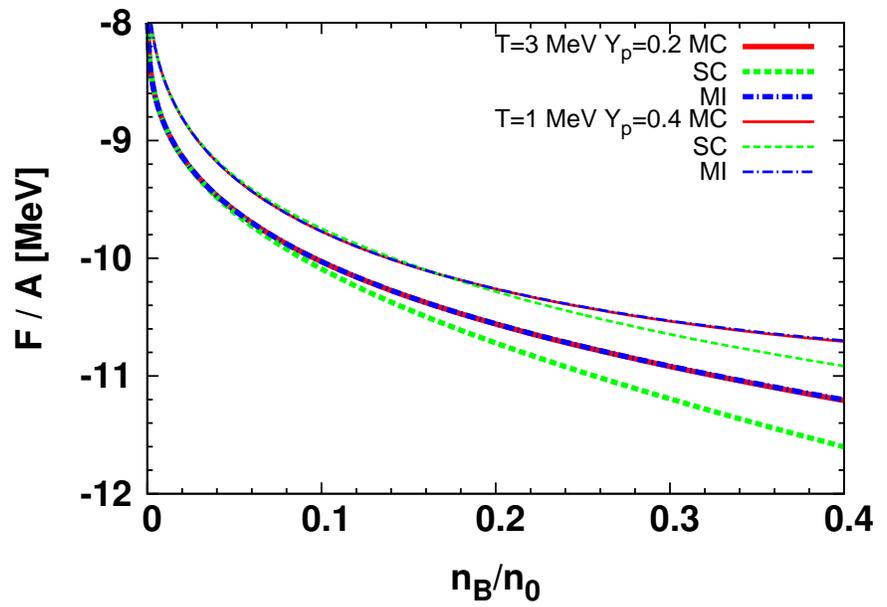}
\caption{Internal free energies per baryon,
$(F^B_i+F^S_i+F^C_i)/A_i $,  
of   $^{50}$Ca  at $(T, Y_p)$=~(3 MeV, 0.2)   (thick lines)
and of $^{300}$Fm at $(T, Y_p)$=~(1 MeV, 0.4) (thin lines),
as  functions of  $n_B$
for 
Models MC (red solid lines), MI (blue dashed-dotted lines) and SC (green dashed  lines).}
\label{fig_bind}
\end{figure}

Figure~\ref{fig_bind} shows the sum of the bulk, surface and Coulomb free energies per baryon, the absolute value of  which corresponds to the binding energy per baryon, for
$^{50}$Ca or   $^{300}$Fm nuclei. Model MC has lower energy than Model MI,
due to the compression or decompression. 
Although  the equilibrium densities of nuclei are changed appreciably, as shown in Fig.~\ref{fig_satuden}, the changes in binding  energies between compressible and incompressible models are rather small.
This is because the change of equilibrium density brings not only the reduction of  some of bulk, surface and Coulomb free energies  but also the increase of  the others as shown in Fig.~\ref{fig_bindshift}.
On the other hand,  the binding energy in Model SC is much larger than in  other MNA  models,
since  more free nucleons reduce the surface free energy.

\begin{figure}
\includegraphics[width=8.1cm]{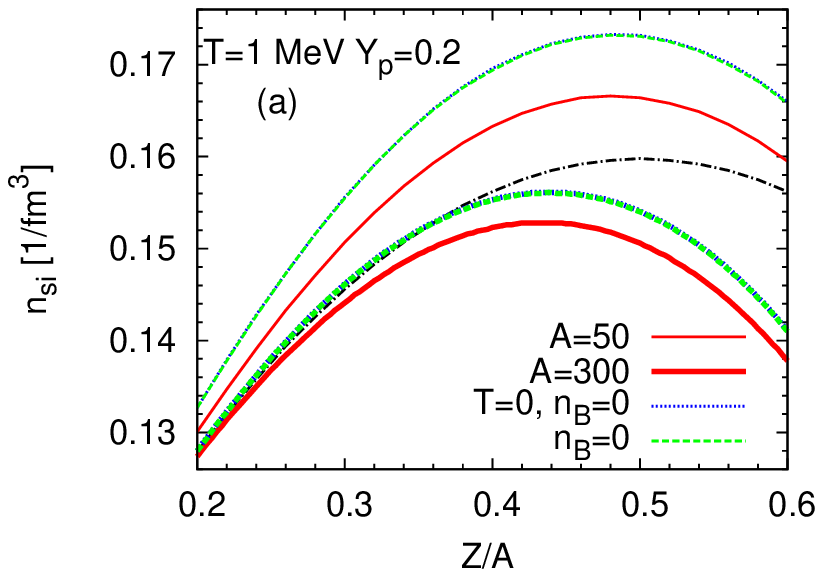}
\includegraphics[width=8.1cm]{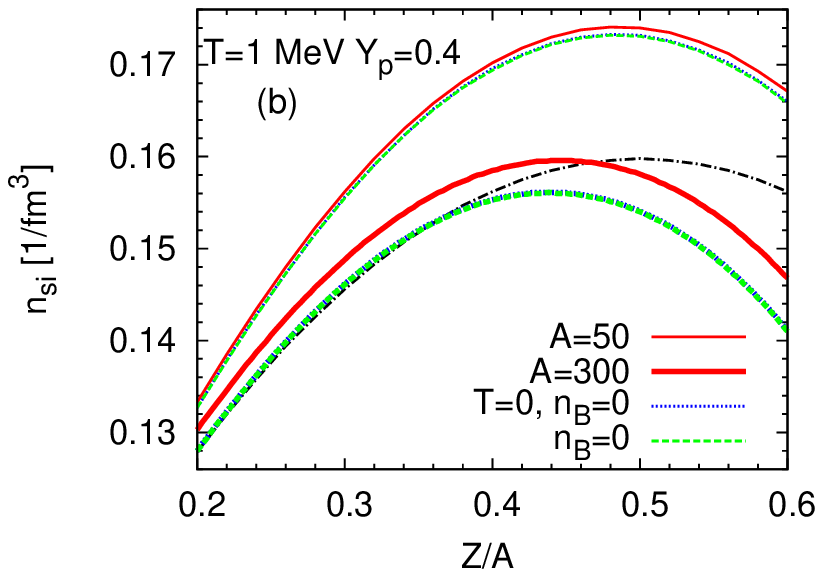}
\includegraphics[width=8.1cm]{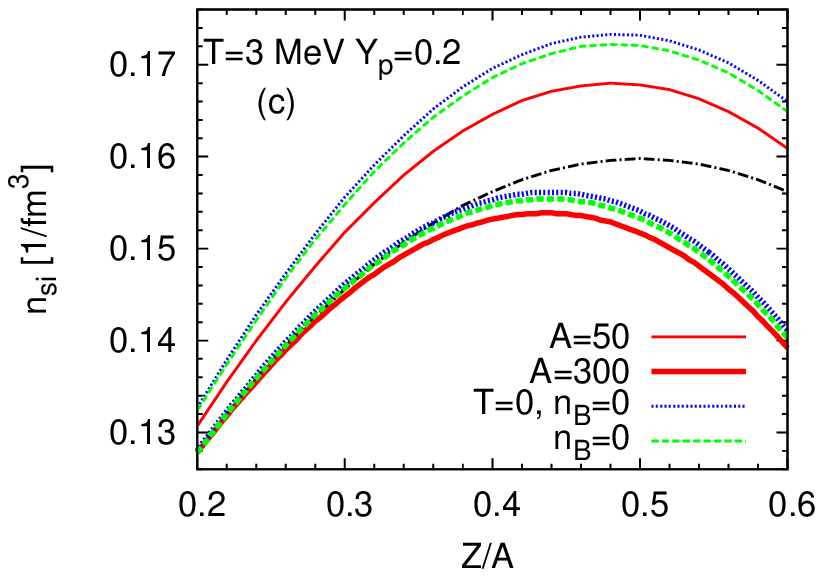}
\includegraphics[width=8.1cm]{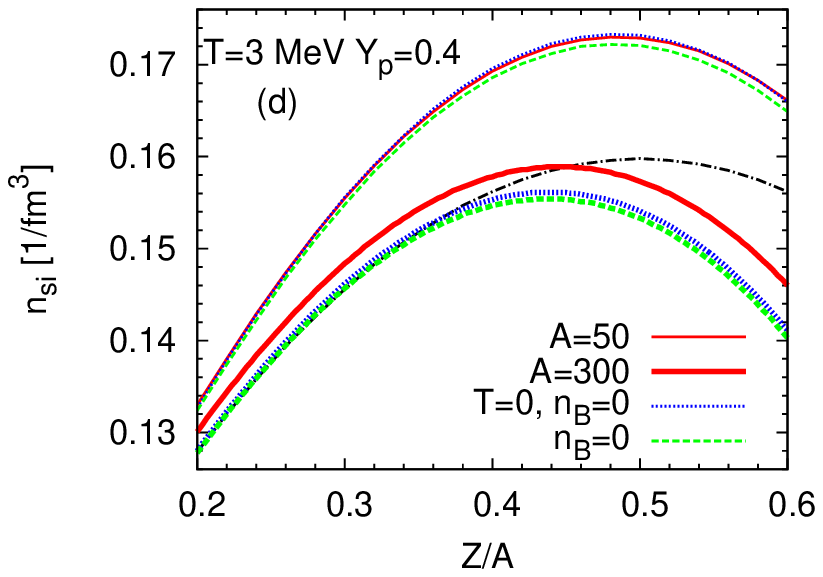}
\caption{Equilibrium densities of the nuclei with $A=$~ 50 (thin)
and  $300$ (thick) 
as a function of proton fraction in nuclei, $Z/A$,  
at $n_B=0.3$~$n_0$ for   
  $T=1$ MeV (top row) and 3 MeV (bottom row)
and $Y_p=$ 0.2 (left column) and 0.4 (right column). 
 Red solid  lines are those obtained in Model MC, $n_{si}(T,n_e,n'_p,n'_n)$. 
Blue dotted and green dashed lines indicate
those in vacuum (or in Model MI), $n_{si}(0,0,0,0)$, and those at finite temperature and zero density, $n_{si}(T,0,0,0)$, respectively.
Black dashed-dotted lines are the density corresponding to the minimum of bulk  free energy $\omega$ as a function of charge fraction $x=Z/A$, $n_{s,bulk}(Z/A)$.
}
\label{fig_satuyp}
\end{figure}

Figure~\ref{fig_satuyp} displays the equilibrium densities 
 of  some nuclei included in the ensemble of heavy nuclei for Model MC at $n_B=0.3 \ n_{0}$,
those in vacuum, $n_{si}(T,n_e,n'_p,n'_n)=n_{si}(0,0,0,0)$, and those at finite temperature  and zero density, $n_{si}(T,0,0,0)$. 
The latter two have almost the same values at $T=1$ MeV, whereas the finite temperature modification of surface energies reduces the equilibrium densities  notably at $T=3$ MeV. 
The nuclei with larger mass numbers have smaller equilibrium densities due to the larger  Coulomb repulsion. 
 Also, the neutron-rich nuclei have smaller equilibrium densities, because the additional symmetry energy  shifts the minimum point of bulk energy, $n_{s,bulk}(x)$, to smaller densities. 
This effect is mainly controlled by the parameter $L$,
which is responsible for the density dependence of the symmetry energy, see Eq.~(\ref{eq:nsbulk}).
We also find that  the nuclei with larger  mass numbers  are likely to be compressed  
for  $Y_p=0.4$,
 since  they have large Coulomb energies. 
As for  neutron-rich matter with $Y_p$=  0.2, the  nuclei with smaller mass numbers  and/or  $Z_i/A_i\sim0.5$  are  more strongly decompressed because of the larger  surface energies.

\begin{figure}
\includegraphics[width=8.1cm]{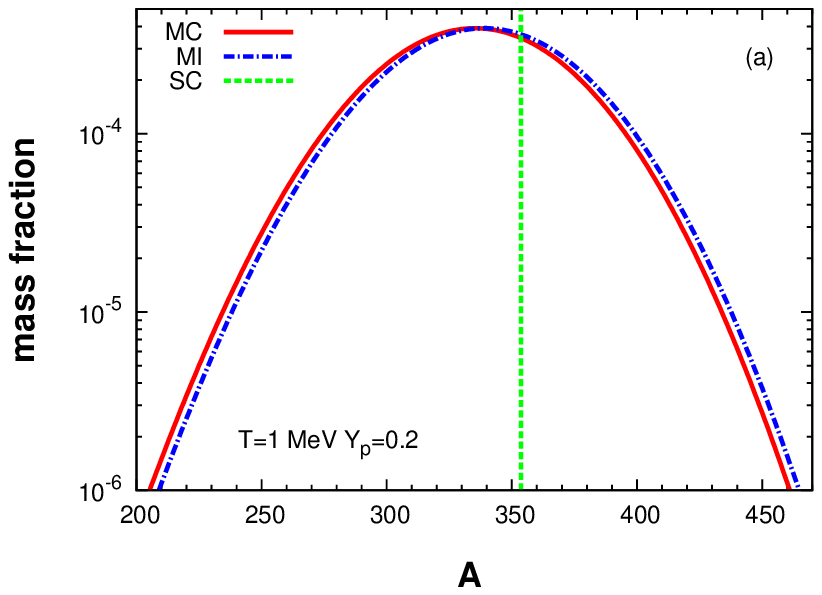}
\includegraphics[width=8.1cm]{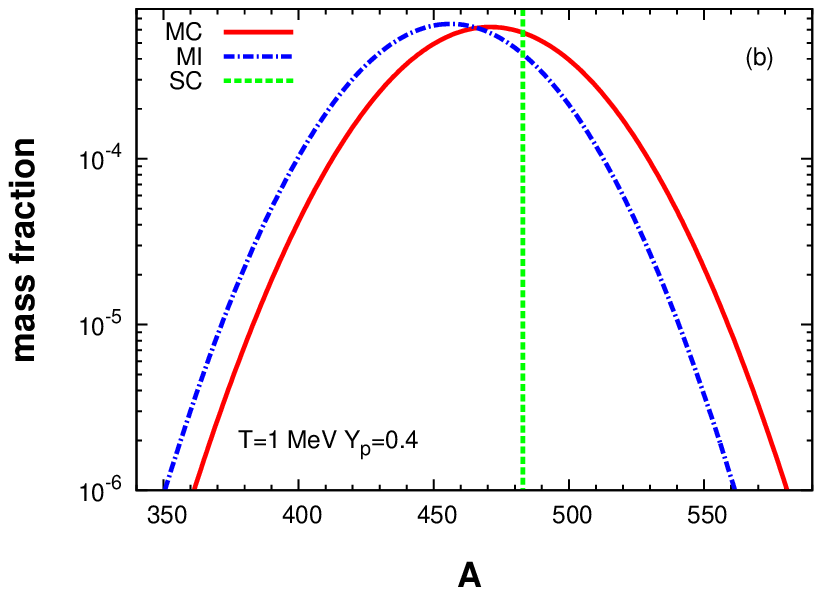}
\includegraphics[width=8.1cm]{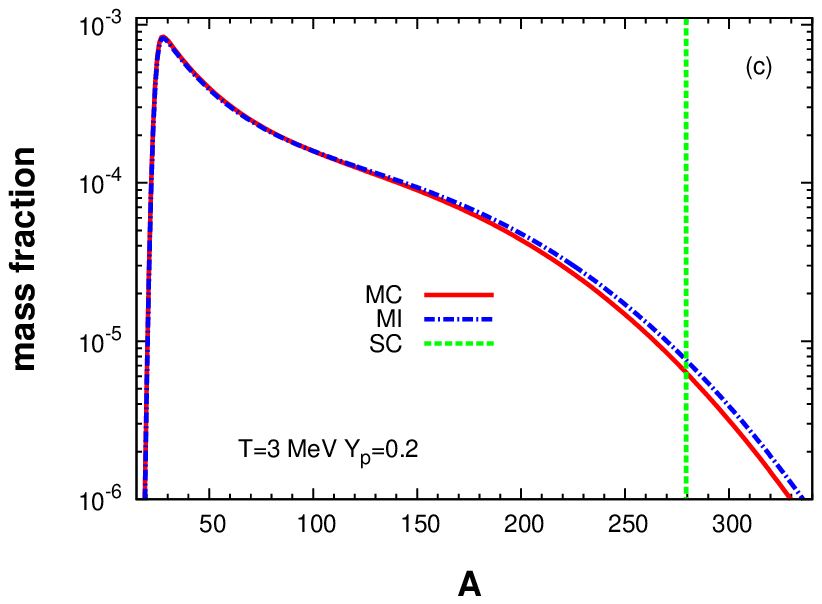}
\includegraphics[width=8.1cm]{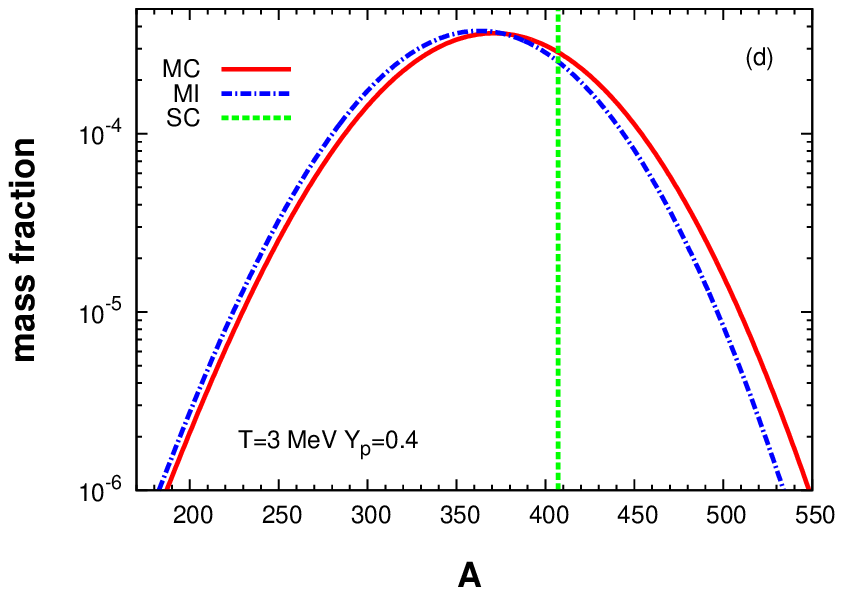}
\caption{
Mass fractions of  elements as a function of mass number
for Models MC (red solid lines) and   MI (blue dashed-dotted lines)  in multi-nucleus description
at  $T=1$ MeV (top row) and 3 MeV (bottom row)
and $Y_p=$ 0.2 (left column) and 0.4 (right column).
Vertical green dashed  lines display the mass numbers of representative nuclei in single nucleus description.
}
\label{fig_massdiss}
\end{figure}

The mass fractions of elements as a  function of  the mass number at $n_B= 0.3$~$n_0$
are  displayed in Fig.~\ref{fig_massdiss}. 
For  $Y_p=0.4$, the Model MC shows the larger mass fraction of nuclei with large mass numbers than Model MI,
 since they give larger compressions, as shown in Fig,~\ref{fig_satuyp}.
On the other hand,  for  $Y_p=0.2$,
the nuclei  with smaller mass numbers are more decompressed and more abundant in  Model MC. 
As a result, Model MC  predicts  the nuclei with smaller mass numbers compared to Model MI.  

\begin{figure}
\includegraphics[width=8.1cm]{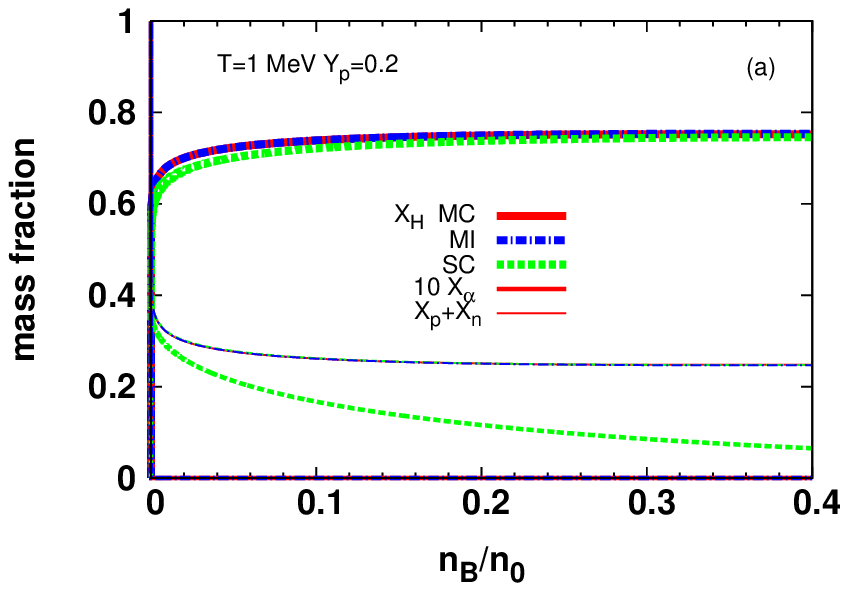}
\includegraphics[width=8.1cm]{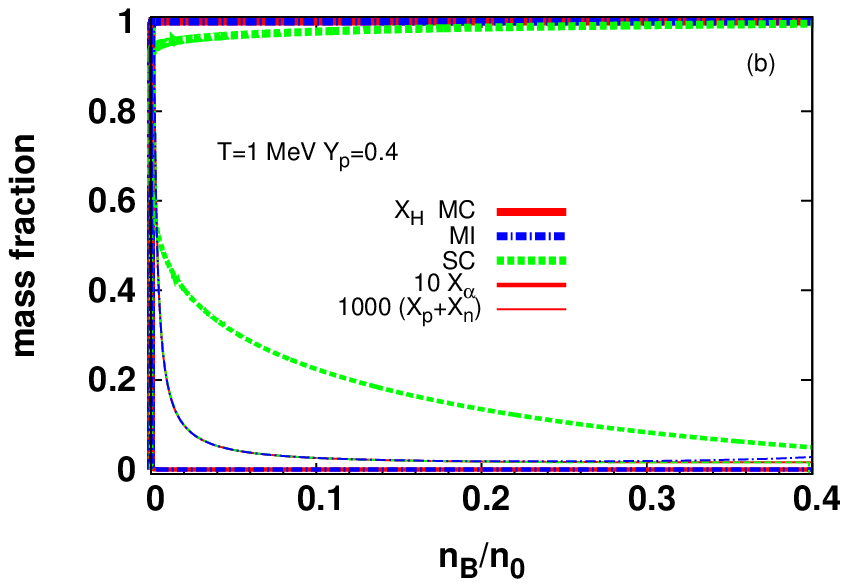}
\includegraphics[width=8.1cm]{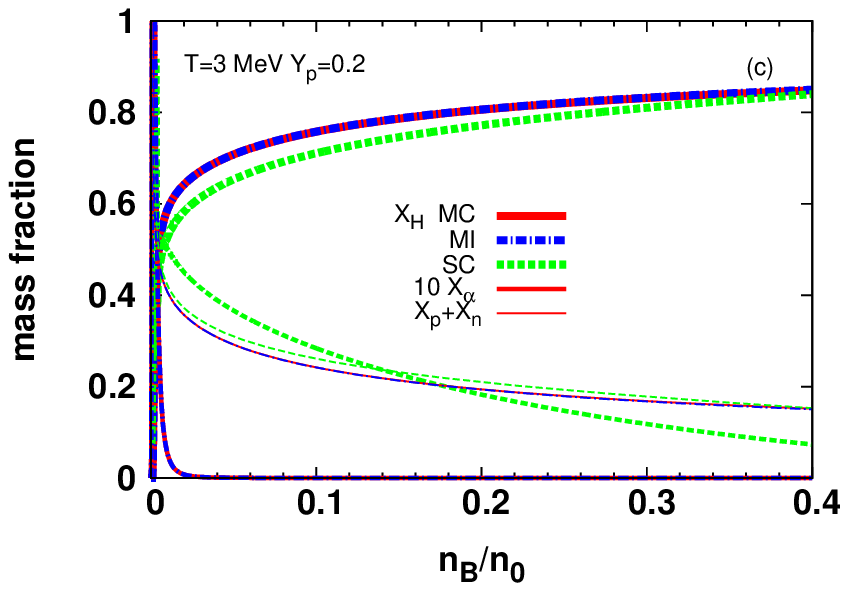}
\includegraphics[width=8.1cm]{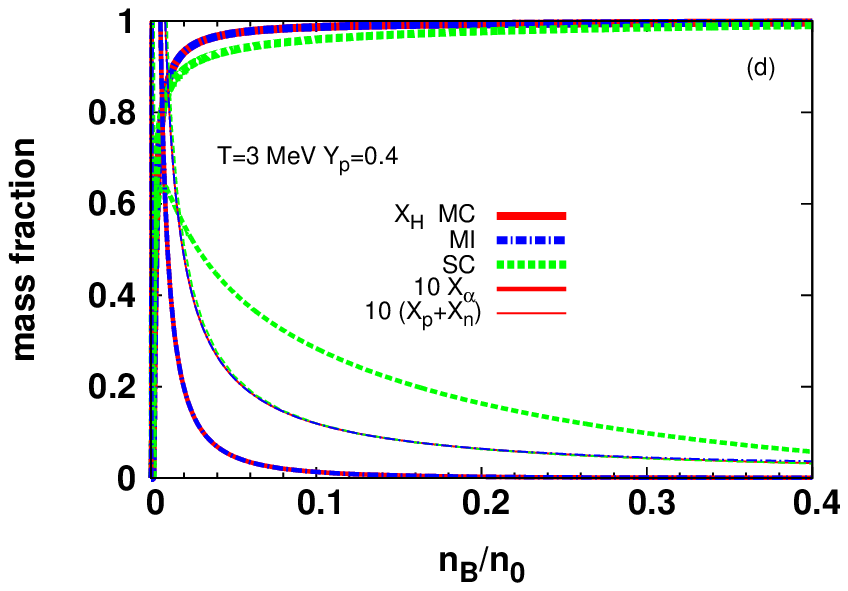}
\caption{Mass fractions of  heavy nuclei (thick  lines),   alpha particles (middle lines) and free nucleons (thin lines)   for Models MC  (red solid lines), SC (green dashed lines) and MI   (blue dashed-dotted lines), respectively,
as a function of  $n_B$
at  $T=1$ MeV (top row) and 3 MeV (bottom row)
and $Y_p=$ 0.2 (left column) and 0.4 (right column).
Those of alpha particles and/or free nucleons are scaled for best display.
}
\label{fig_massfrac}
\end{figure}

\begin{figure}
\includegraphics[width=8.1cm]{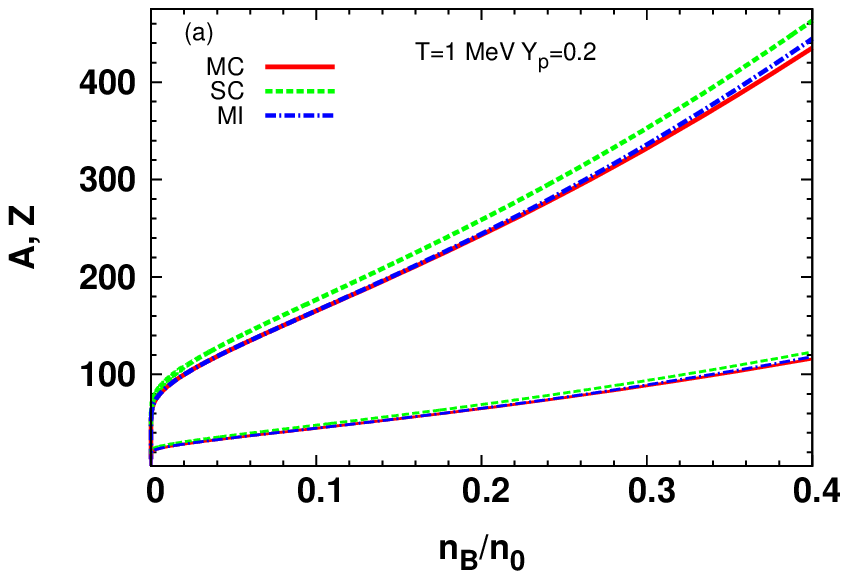}
\includegraphics[width=8.1cm]{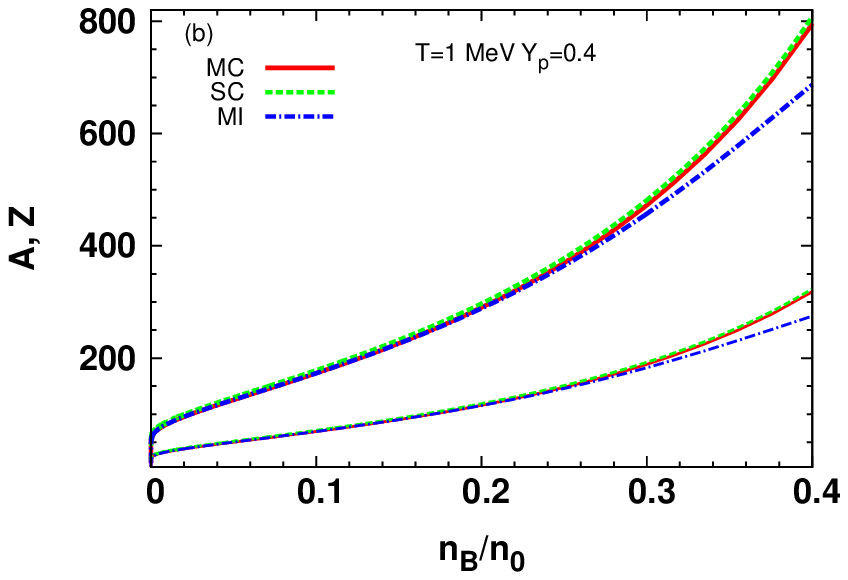}
\includegraphics[width=8.1cm]{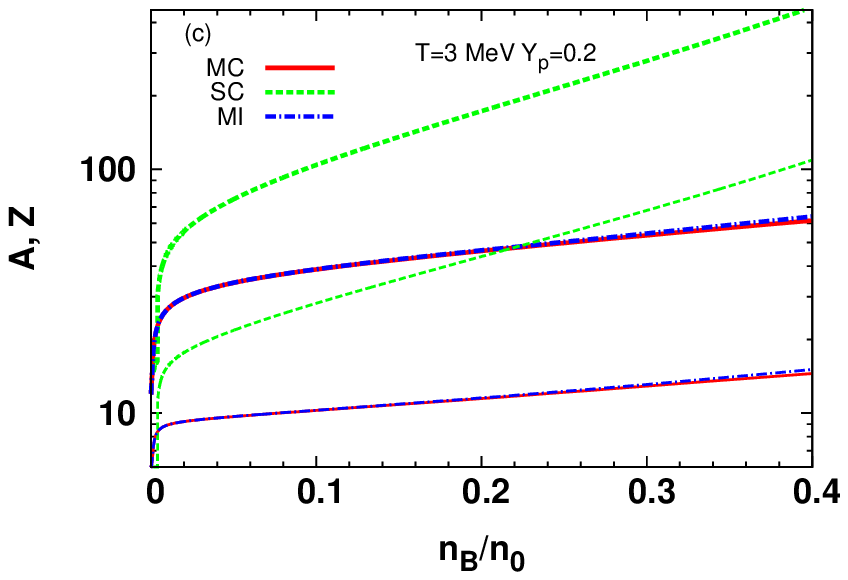}
\includegraphics[width=8.1cm]{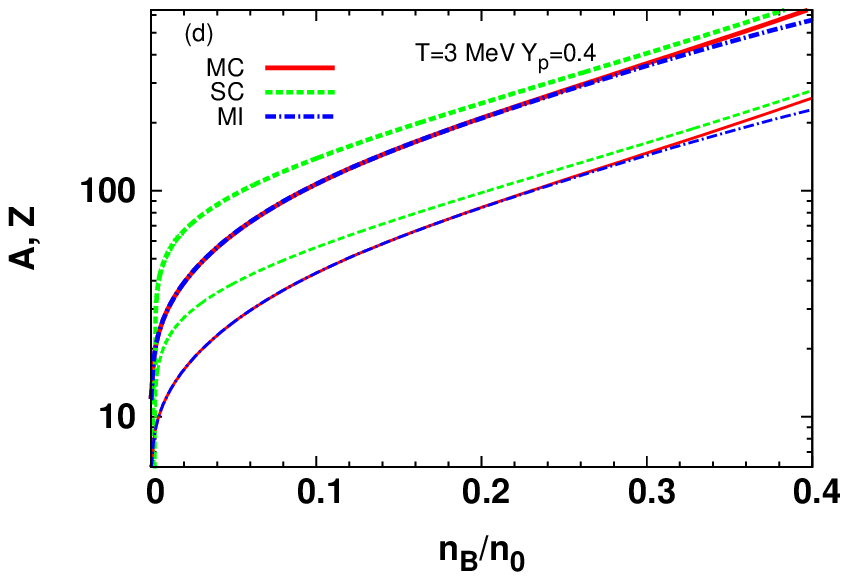}
\caption{Average mass  and proton numbers   (thick and thin lines) of heavy nuclei with $Z \geq 6 $ 
for   Models MC  (red solid lines) and MI  (blue dashed-dotted lines)
 and  those of representative nucleus for  Model  SC (green dashed lines)
 as a function of  $n_B$
at  $T=1$ MeV (top row) and 3 MeV (bottom row)
and $Y_p=$~0.2 (left column) and 0.4 (right column).
}
\label{fig_massnumber}
\end{figure}

 The  mass fraction of  free nucleons, alpha particles and heavy nuclei ($Z_i \geq 6$)
and  average mass and proton numbers of heavy nuclei  are displayed in Figs.~\ref{fig_massfrac} and~\ref{fig_massnumber}, respectively, %
 as functions of baryon density.
The total mass fraction of heavy nuclei in model MC is always  larger  than in models MI and SC,  
because  of additional degrees of freedom in the description of heavy nuclei.
 We also find that the mass fractions of alpha particles are significantly overestimated in Model SC.
For  $Y_p$=~ 0.4, the Model MC shows the larger mass numbers of heavy nuclei than Model MI,
because the formation of such nuclei is favored  by
 the compression, as shown in Fig.~\ref{fig_massdiss}.
On the other hand,
for  $Y_p$= 0.2,
the nuclei  with smaller mass numbers are more abundant in  Model MC that leads to the  smaller average mass number than that in Model MI.  
  Model SC  gives larger mass numbers than multi-nucleus models,  as in Ref. \cite{burrows84}.
The deviations of Model SC from Model MC  increase with  temperature,
because nuclei other than representative nucleus  become more and more abundant but the mass fraction of the peak nuclei is reduced.

\begin{figure}
\includegraphics[width=12cm]{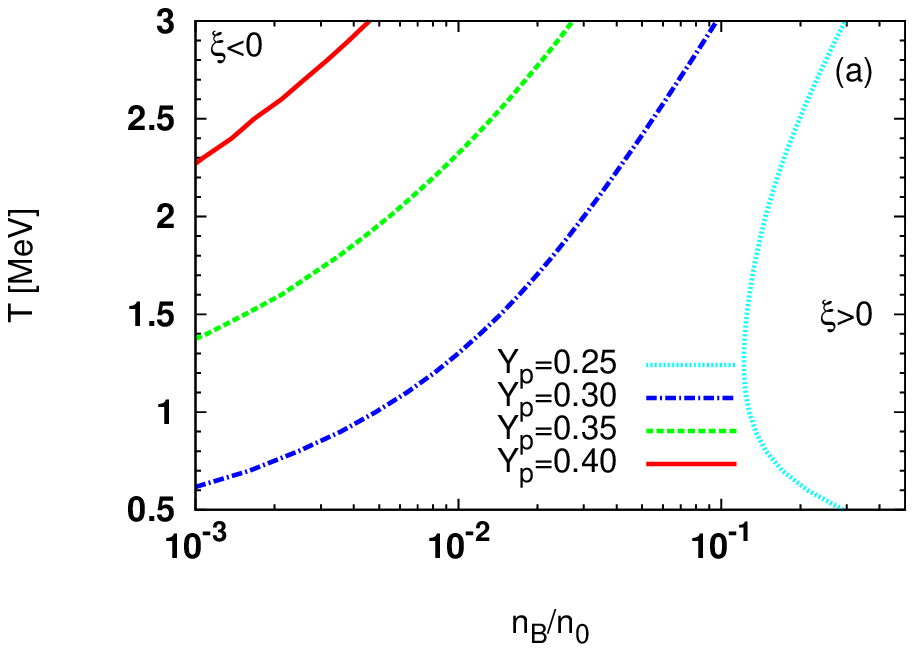}
\includegraphics[width=12cm]{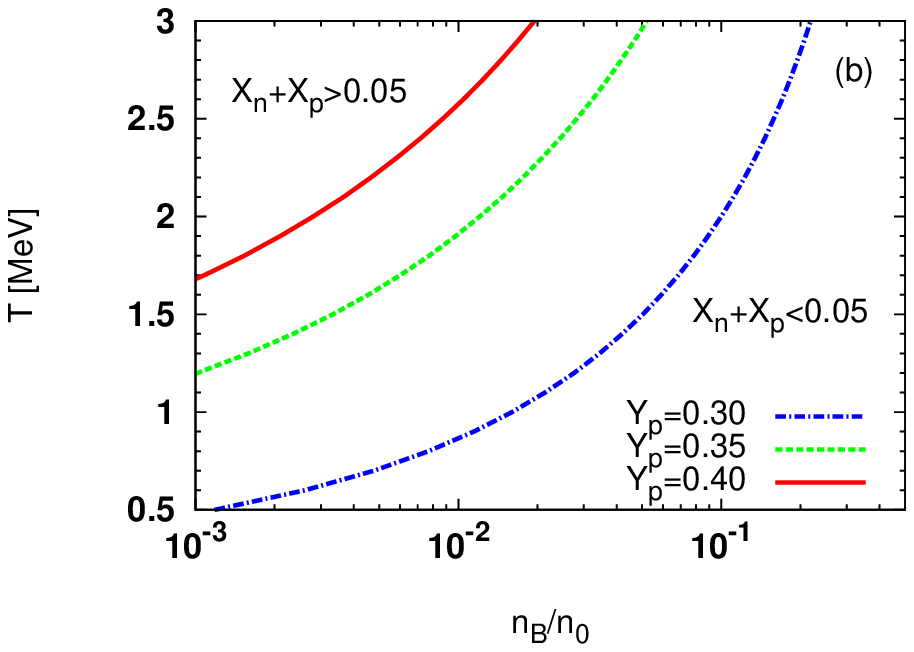}
\caption{Critical lines  for the  compression (top)
and mass fraction of free nucleons  (bottom)
in Model MC
for $Y_p=0.25$ (cyan dotted line), 0.30 (blue dashed-dotted line), 0.35 (green  dashed line) and  0.40 (red  solid line). 
They are defined as the  compression parameter 
 $\xi = \sum_i n_i ({\rm d}  n_{si}/ {\rm d} n_{B})/(\sum_i n_i)=0$
and the mass fraction of free nucleons $X_p+X_n=0.05$. 
The positive value of $\xi$ means that in general the nuclei are compressed.
Note that the line of mass fraction for $Y_p =0.25$ is not displayed, since  $X_p+X_n > 0.05$ at all conditions.
} 
\label{fig_phase}
\end{figure}

Figure~\ref{fig_phase}a shows the lines
in $n_B-T$ plane separating regions where
the nuclei are compressed or decompressed 
that are determined by the
 compression parameter    $\xi = \sum_i n_i ({\rm d}  n_{si}/ {\rm d} n_{B})/(\sum_i n_i)$.
The positive value of $\xi$ means that, on average,  the nuclei are compressed, 
 and the negative one  corresponds to the decompression.
The critical lines of the mass fraction of free nucleons  $X_n+X_p=0.05$
are also shown in Fig.~\ref{fig_phase}b.
Note that neutrons are always dripped and   $X_n \gtrsim 0.05$ at $Y_p \lesssim0.25$.
Free nucleons are almost diminished and nuclei are dominant at high densities, low temperatures and high proton fractions.  In those states, the nuclei are compressed due to the reduced Coulomb energies.
We find that the  compression is replaced by decompression  when 
 the mass fraction of  free nucleons $X_n+X_p$ becomes less than $\sim 0.05-0.15$.
In core-collpase supernovae and neutron star  mergers,  such thermodynamical  conditions, $n_B \gtrsim 0.01 \  n_0$ and $Y_p \gtrsim 0.25$, are usually not realized because of de-leptonization \cite{sullivan16,sekiguchi15,sekiguchi16}. 
Therefore, the nuclei are  likely to be decompressed due to the  presence of dripped neutrons under these conditions. 
However, during  core-collapse of massive stars,  the nuclei may be   compressed due to the surrounding electrons  just before the bounce at
$n_B \gtrsim 0.3 n_0$, 
$T\sim3$ MeV and  $Y_p \sim 0.25$. 
The parameter $\xi$ is zero in Model MI, and negative  in Model SC
at all densities.

\begin{figure}
\includegraphics[width=8.1cm]{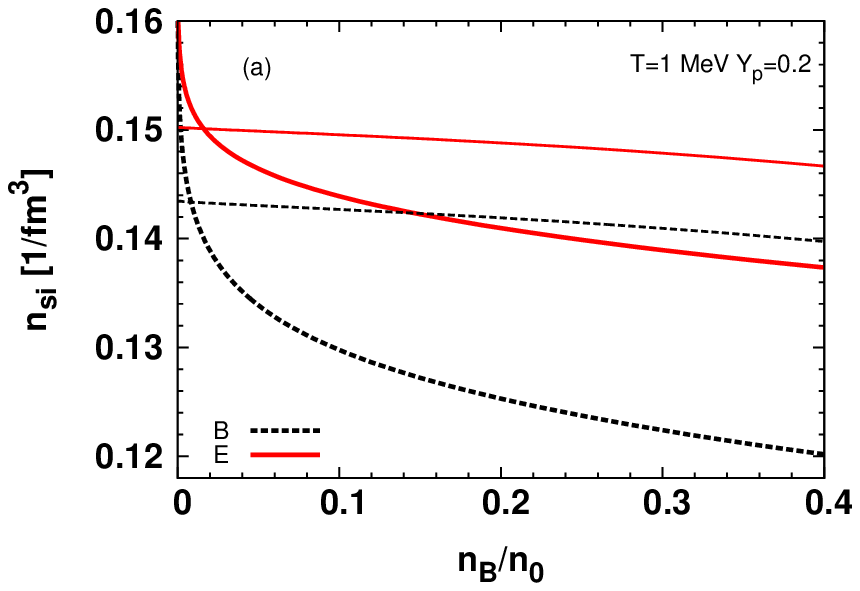}
\includegraphics[width=8.1cm]{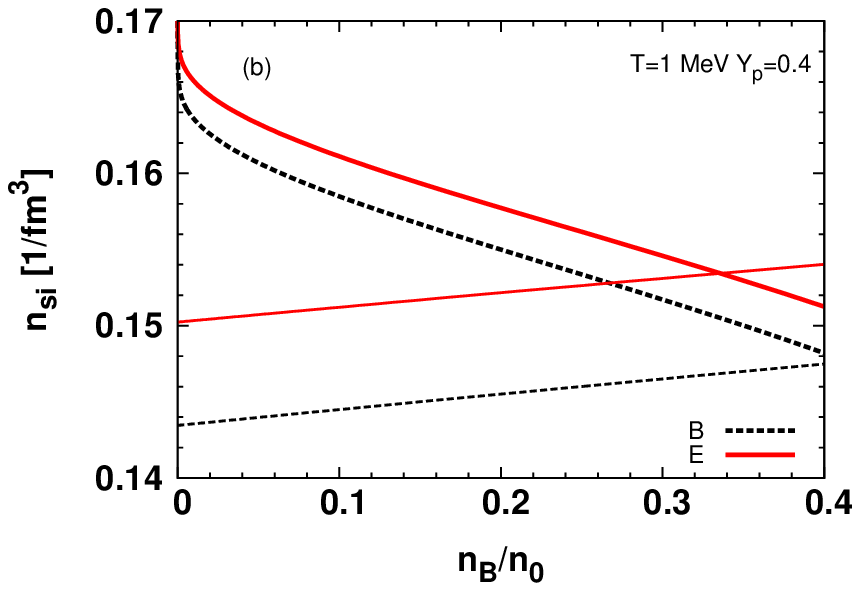}
\includegraphics[width=8.1cm]{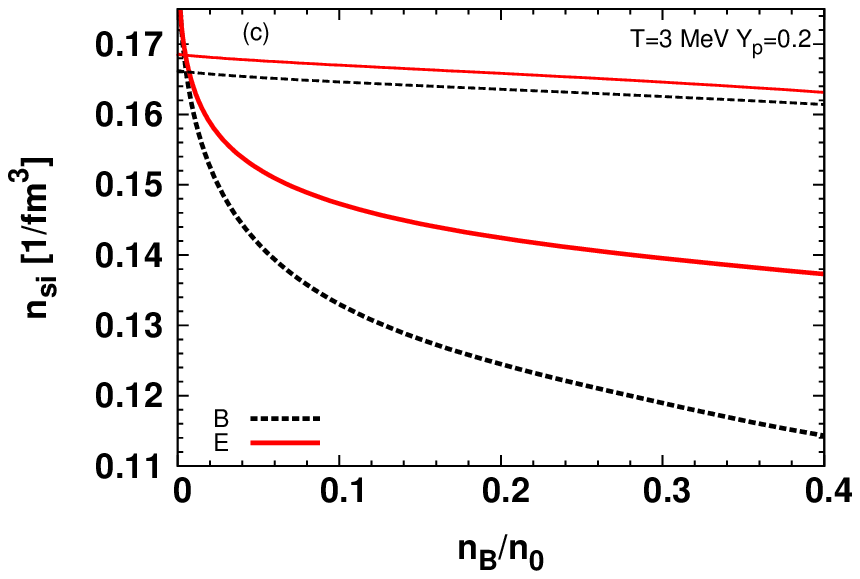}
\includegraphics[width=8.1cm]{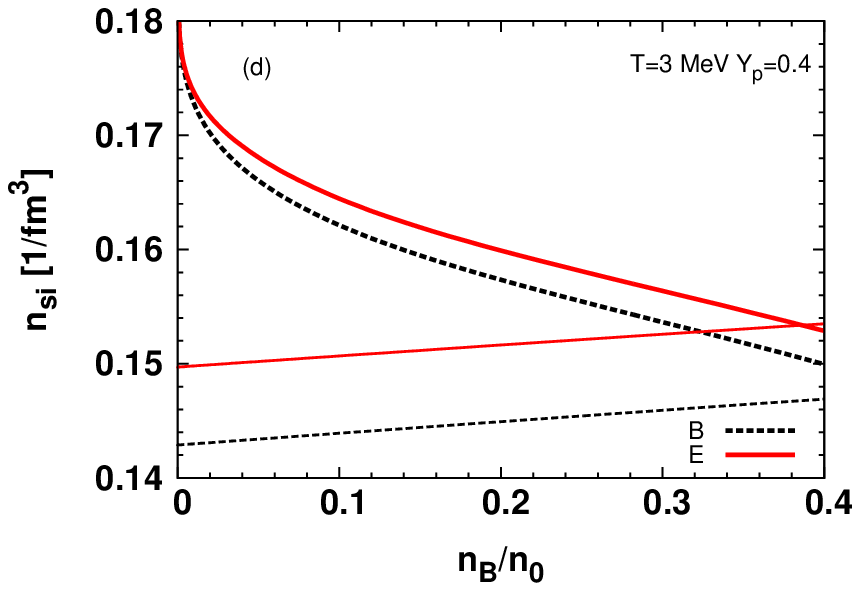}
\caption{Average  equilibrium density (thick lines) of heavy nuclei  as a function of  $n_B$
at  $T=1$ MeV (top row) and 3 MeV (bottom row)
and $Y_p=$ 0.2 (left column) and 0.4 (right column)
for  MC models  with bulk parameter set B (black dashed lines)  and E (red thick lines),
the latter of which is identical to the MC model shown in Fig.~\ref{fig_satuden}.
Thin lines indicate  those of
$^{50}_{20}$Ca at $(T, Y_p)$=~(3 MeV, 0.2)
and   $^{300}_{100}$Fm at the other conditions.
}
\label{fig_satudenl}
\end{figure}

\begin{figure}
\includegraphics[width=12cm]{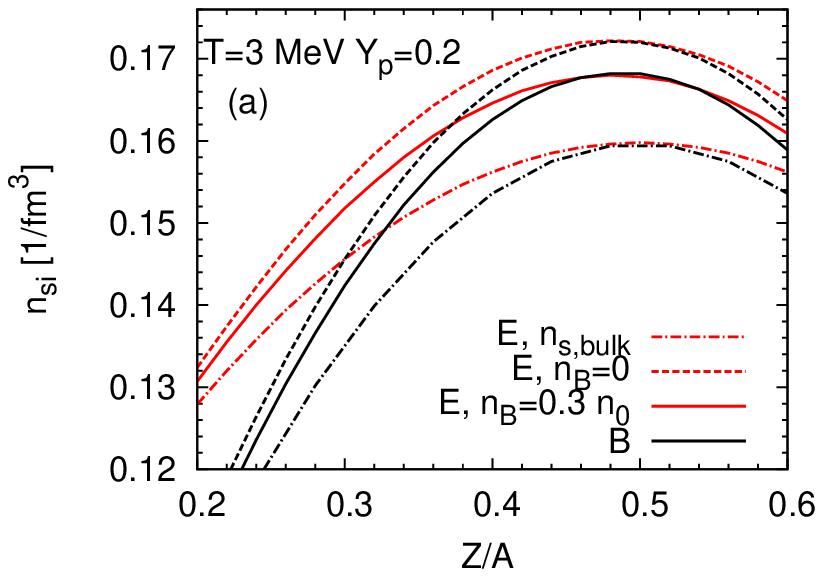}
\includegraphics[width=12cm]{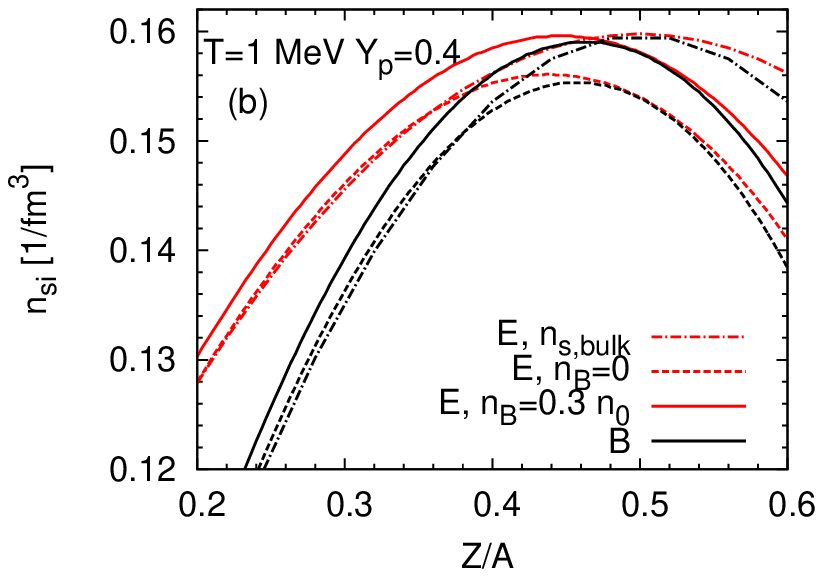}
\caption{Equilibrium densities of the nuclei with $A=$~ 50 (thin) at $(T, Y_p)$=~(3 MeV, 0.2)  (top panel)
and  $300$ (thick) at $(T, Y_p)$=~(1 MeV, 0.4) (bottom panel)
as a function of proton fraction in nuclei, $Z/A$,
for   MC models  with  bulk parameter set B (black lines)  and E (red lines).
Solid, dashed and dashed-dotted  lines are those obtained at $n_B=0.3$~$n_0$, those   at finite temperature and zero density,
and the density   corresponding to the minimum of bulk  free energy, $n_{s,bulk}(Z/A)$, respectively.
}
\label{fig_satuypl}
\end{figure}

Finally, we discuss the impact of  bulk properties on these results by using another bulk parameter set B  presented in Tab.~\ref{tab1_bulk}.
The major difference between the  parameter sets B and E in the bulk properties is the slope parameter $L$, whereas the same $K$ is assumed. 
The other parameters $S_0$, $\omega_0$, $n_0$ are slightly modified to reproduce properties of stable nuclei \cite{oyamatsu03}.
Figs.~\ref{fig_satudenl} and \ref{fig_satuypl} display the average and individual equilibrium densities
  for Models MC based on these parameter sets, in which 
the results for the parameter set E  are identical to those in
 Figs.~\ref{fig_satuden} and \ref{fig_satuyp}.
 Both average and individual equilibrium densities in the model of the parameter set B with larger $L$ are considerably smaller than those in the reference model  of the parameter set E, as we expected. This is because the neutron-rich bulk matter has smaller saturation  density as described by Eq.~(\ref{eq:nsbulk}), thereby resulting in the smaller equilibrium density of  neutron-rich nuclei. On the other hand, the qualitative features in the equilibration of individual nuclei are not sensitive to the $L$ parameter; the nuclei are compressed and decompressed for $Y_p=$0.2 and 0.4, respectively.

\section{Conclusion \label{sec:conc}}
We have constructed the multi-nucleus EOS based on the compressible liquid drop model for the  stellar matter at sub-nuclear densities and finite temperatures.
The equilibrium densities of individual  nuclei are calculated self consistently.
We have compared
 the single- and multi-nucleus descriptions, in which the free energies are minimized for nuclear structure and for nuclear ensemble, respectively. 
We have found that in-medium nuclei are compressed   because   Coulomb energies are reduced by background electrons under the condition 
that few vapor nucleons are  present  ($X_n+X_p \lesssim 0.05-0.15$), e.g. at $Y_p = 0.4$. But dripped nucleons reduce the surface free energies and induce  decompressions of nuclei at the neutron-rich condition such as $Y_p = 0.2$.  
The average equilibrium density decreases as the baryon density increases regardless of whether equilibrium densities of individual  nuclei  are compressed or decompressed, since the nuclei with larger mass numbers and smaller equilibrium densities dominate in the ensemble.
We have found that the property
 of nuclei are affected  to some degree by the compression or decompression even  at  low baryon densities around $n_B \sim 0.01 \ n_0$,
 although the noticeable changes in binding energies, total mass fractions and average mass numbers are found only at density above $0.1 \ n_0$. 
In the core-collapse supernova and neutron star mergers, 
the nuclei are likely to be decompressed due to dripped neutrons,
 except just before the core bounce of massive stars. 
We 
come to the conclusion that the equilibrium densities of nuclei used in  some single-nucleus EOS calculations, see e.g. \cite{shen11,papakonstantinou13,aymard14}, 
 may be significantly underestimated
due to the incorrect density of vapor nucleons.
This means that the SNA may not be realistic enough to derive reliable information  
 about in-medium effects as a function of $n_B, T, Y_p$.

Our model is not perfect because it neglects nuclear forces and Pauli blocking among free nucleons, nuclear shell effects and  light clusters other than the alpha particle.
The Pauli shifts for neutrons would suppress the abundance of nucleons and the impact of surface energy reduction by the neutron vapor may become weak, although the impacts are not so large due to the small neutron fraction for the cases of  $Y_p=0.4$ and/or $T=3$ MeV.
The shell effect should be included and may be affected in medium  as well as bulk properties,  especially at low temperatures, whereas they are  completely washed out at $T \sim 3$ MeV and negligible. 
The existence of deuterons and  tritons is known to reduce the neutron fraction a little \cite{sumiyoshi08}, thereby suppressing the surface modification by neutrons.
In addition, we should also consider the nuclear pasta phases which are formed above $n_B \gtrsim 0.3 n_0$ yet neglected in this study.
As a next step, we plan to address deformations of nuclei including those nuclear pasta phases in MNA.
Note that  there remain  uncertainties in estimation of surface energies such as the formulation of its dependence on the nucleon vapor density and choice of $\sigma_0$. They also may affect the results  quantitatively to some degree. 
Nevertheless, we hope that our study will provide the helpful information  for
further improvement of the nuclear EOS required for  SN simulations.
These in-medium effects may lead to significant changes of the dynamics and  weak reaction rates in the core-collapse supernovae and neutron star mergers.

\begin{acknowledgments}
S.F is thankful to  S. Yamada, K. Sumiyoshi and H. Suzuki for the discussion about FYSS EOS which made  the basis of this study.
I. M thanks S.Typel and M. Hempel for useful discussions
regarding in-medium modifications of nuclei. 
S.F.  is supported by Japan Society for the Promotion of
Science Postdoctoral Fellowships for Research Abroad. 
A part of the numerical calculations were carried out on  PC cluster at Center
for Computational Astrophysics, National Astronomical Observatory of Japan.
\end{acknowledgments}

\bibliography{reference170128}

\begin{thebibliography}{56}
\expandafter\ifx\csname natexlab\endcsname\relax\def\natexlab#1{#1}\fi
\expandafter\ifx\csname bibnamefont\endcsname\relax
  \def\bibnamefont#1{#1}\fi
\expandafter\ifx\csname bibfnamefont\endcsname\relax
  \def\bibfnamefont#1{#1}\fi
\expandafter\ifx\csname citenamefont\endcsname\relax
  \def\citenamefont#1{#1}\fi
\expandafter\ifx\csname url\endcsname\relax
  \def\url#1{\texttt{#1}}\fi
\expandafter\ifx\csname urlprefix\endcsname\relax\def\urlprefix{URL }\fi
\providecommand{\bibinfo}[2]{#2}
\providecommand{\eprint}[2][]{\url{#2}}

\bibitem[{\citenamefont{{Janka}}(2012)}]{janka12}
\bibinfo{author}{\bibfnamefont{H.-T.} \bibnamefont{{Janka}}},
  \bibinfo{journal}{Annual Review of Nuclear and Particle Science}
  \textbf{\bibinfo{volume}{62}}, \bibinfo{pages}{407} (\bibinfo{year}{2012}). 

\bibitem[{\citenamefont{{Kotake} et~al.}(2012)\citenamefont{{Kotake},
  {Takiwaki}, {Suwa}, {Iwakami Nakano}, {Kawagoe}, {Masada}, and
  {Fujimoto}}}]{kotake12}
\bibinfo{author}{\bibfnamefont{K.}~\bibnamefont{{Kotake}}},
  \bibinfo{author}{\bibfnamefont{T.}~\bibnamefont{{Takiwaki}}},
  \bibinfo{author}{\bibfnamefont{Y.}~\bibnamefont{{Suwa}}},
  \bibinfo{author}{\bibfnamefont{W.}~\bibnamefont{{Iwakami Nakano}}},
  \bibinfo{author}{\bibfnamefont{S.}~\bibnamefont{{Kawagoe}}},
  \bibinfo{author}{\bibfnamefont{Y.}~\bibnamefont{{Masada}}}, \bibnamefont{and}
  \bibinfo{author}{\bibfnamefont{S.-i.} \bibnamefont{{Fujimoto}}},
  \bibinfo{journal}{Advances in Astronomy} \textbf{\bibinfo{volume}{2012}},
  \bibinfo{eid}{428757} (\bibinfo{year}{2012}).

\bibitem[{\citenamefont{{Burrows}}(2013)}]{burrows13}
\bibinfo{author}{\bibfnamefont{A.}~\bibnamefont{{Burrows}}},
  \bibinfo{journal}{Reviews of Modern Physics} \textbf{\bibinfo{volume}{85}},
  \bibinfo{pages}{245} (\bibinfo{year}{2013}).

\bibitem[{\citenamefont{{Foglizzo} et~al.}(2015)\citenamefont{{Foglizzo},
  {Kazeroni}, {Guilet}, {Masset}, {Gonz{\'a}lez}, {Krueger}, {Novak}, {Oertel},
  {Margueron}, {Faure} et~al.}}]{foglizzo15}
\bibinfo{author}{\bibfnamefont{T.}~\bibnamefont{{Foglizzo}}},
  \bibinfo{author}{\bibfnamefont{R.}~\bibnamefont{{Kazeroni}}},
  \bibinfo{author}{\bibfnamefont{J.}~\bibnamefont{{Guilet}}},
  \bibinfo{author}{\bibfnamefont{F.}~\bibnamefont{{Masset}}},
  \bibinfo{author}{\bibfnamefont{M.}~\bibnamefont{{Gonz{\'a}lez}}},
  \bibinfo{author}{\bibfnamefont{B.~K.} \bibnamefont{{Krueger}}},
  \bibinfo{author}{\bibfnamefont{J.}~\bibnamefont{{Novak}}},
  \bibinfo{author}{\bibfnamefont{M.}~\bibnamefont{{Oertel}}},
  \bibinfo{author}{\bibfnamefont{J.}~\bibnamefont{{Margueron}}},
  \bibinfo{author}{\bibfnamefont{J.}~\bibnamefont{{Faure}}},
  \bibnamefont{et~al.}, \bibinfo{journal}{Publications of the Astronomical
  Society of Australia} \textbf{\bibinfo{volume}{32}}, \bibinfo{eid}{e009}
  (\bibinfo{year}{2015}).

\bibitem[{\citenamefont{Shibata and Taniguchi}(2011)}]{shibata11}
\bibinfo{author}{\bibfnamefont{M.}~\bibnamefont{Shibata}} \bibnamefont{and}
  \bibinfo{author}{\bibfnamefont{K.}~\bibnamefont{Taniguchi}},
  \bibinfo{journal}{Living Reviews in Relativity} \textbf{\bibinfo{volume}{14}}
  (\bibinfo{year}{2011}). 

\bibitem[{\citenamefont{Faber and Rasio}(2012)}]{faber12}
\bibinfo{author}{\bibfnamefont{J.~A.} \bibnamefont{Faber}} \bibnamefont{and}
  \bibinfo{author}{\bibfnamefont{F.~A.} \bibnamefont{Rasio}},
  \bibinfo{journal}{Living Reviews in Relativity} \textbf{\bibinfo{volume}{15}}
  (\bibinfo{year}{2012}).

\bibitem[{\citenamefont{{Raduta} et~al.}(2016)\citenamefont{{Raduta},
  {Gulminelli}, and {Oertel}}}]{raduta16}
\bibinfo{author}{\bibfnamefont{A.~R.} \bibnamefont{{Raduta}}},
  \bibinfo{author}{\bibfnamefont{F.}~\bibnamefont{{Gulminelli}}},
  \bibnamefont{and} \bibinfo{author}{\bibfnamefont{M.}~\bibnamefont{{Oertel}}},
  \bibinfo{journal}{\prc} \textbf{\bibinfo{volume}{93}}, \bibinfo{eid}{025803}
  (\bibinfo{year}{2016}).

\bibitem[{\citenamefont{{Furusawa} et~al.}(2017)\citenamefont{{Furusawa},
  {Nagakura}, {Sumiyoshi}, {Kato}, and {Yamada}}}]{furusawa17b}
\bibinfo{author}{\bibfnamefont{S.}~\bibnamefont{{Furusawa}}},
  \bibinfo{author}{\bibfnamefont{H.}~\bibnamefont{{Nagakura}}},
  \bibinfo{author}{\bibfnamefont{K.}~\bibnamefont{{Sumiyoshi}}},
  \bibinfo{author}{\bibfnamefont{C.}~\bibnamefont{{Kato}}}, \bibnamefont{and}
  \bibinfo{author}{\bibfnamefont{S.}~\bibnamefont{{Yamada}}},
  \bibinfo{journal}{\prc} \textbf{\bibinfo{volume}{95}}, \bibinfo{eid}{025809}
 (\bibinfo{year}{2017}).

\bibitem[{\citenamefont{{Hix} et~al.}(2003)\citenamefont{{Hix}, {Messer},
  {Mezzacappa}, {Liebend{\"o}rfer}, {Sampaio}, {Langanke}, {Dean}, and
  {Mart{\'{\i}}nez-Pinedo}}}]{hix03}
\bibinfo{author}{\bibfnamefont{W.~R.} \bibnamefont{{Hix}}},
  \bibinfo{author}{\bibfnamefont{O.~E.} \bibnamefont{{Messer}}},
  \bibinfo{author}{\bibfnamefont{A.}~\bibnamefont{{Mezzacappa}}},
  \bibinfo{author}{\bibfnamefont{M.}~\bibnamefont{{Liebend{\"o}rfer}}},
  \bibinfo{author}{\bibfnamefont{J.}~\bibnamefont{{Sampaio}}},
  \bibinfo{author}{\bibfnamefont{K.}~\bibnamefont{{Langanke}}},
  \bibinfo{author}{\bibfnamefont{D.~J.} \bibnamefont{{Dean}}},
  \bibnamefont{and}
  \bibinfo{author}{\bibfnamefont{G.}~\bibnamefont{{Mart{\'{\i}}nez-Pinedo}}},
  \bibinfo{journal}{Physical Review Letters} \textbf{\bibinfo{volume}{91}}.

\bibitem[{\citenamefont{{Lentz} et~al.}(2012)\citenamefont{{Lentz},
  {Mezzacappa}, {Messer}, {Hix}, and {Bruenn}}}]{lentz12}
\bibinfo{author}{\bibfnamefont{E.~J.} \bibnamefont{{Lentz}}},
  \bibinfo{author}{\bibfnamefont{A.}~\bibnamefont{{Mezzacappa}}},
  \bibinfo{author}{\bibfnamefont{O.~E.~B.} \bibnamefont{{Messer}}},
  \bibinfo{author}{\bibfnamefont{W.~R.} \bibnamefont{{Hix}}}, \bibnamefont{and}
  \bibinfo{author}{\bibfnamefont{S.~W.} \bibnamefont{{Bruenn}}},
  \bibinfo{journal}{\apj} \textbf{\bibinfo{volume}{760}}, \bibinfo{eid}{94}
  (\bibinfo{year}{2012}).

\bibitem[{\citenamefont{Wanajo et~al.}(2014)\citenamefont{Wanajo, Sekiguchi,
  Nishimura, Kiuchi, Kyutoku, and Shibata}}]{wanajo14}
\bibinfo{author}{\bibfnamefont{S.}~\bibnamefont{Wanajo}},
  \bibinfo{author}{\bibfnamefont{Y.}~\bibnamefont{Sekiguchi}},
  \bibinfo{author}{\bibfnamefont{N.}~\bibnamefont{Nishimura}},
  \bibinfo{author}{\bibfnamefont{K.}~\bibnamefont{Kiuchi}},
  \bibinfo{author}{\bibfnamefont{K.}~\bibnamefont{Kyutoku}}, \bibnamefont{and}
  \bibinfo{author}{\bibfnamefont{M.}~\bibnamefont{Shibata}},
  \bibinfo{journal}{The Astrophysical Journal Letters}
  \textbf{\bibinfo{volume}{789}}, \bibinfo{pages}{L39} (\bibinfo{year}{2014}). 

\bibitem[{\citenamefont{Sekiguchi et~al.}(2015)\citenamefont{Sekiguchi, Kiuchi,
  Kyutoku, and Shibata}}]{sekiguchi15}
\bibinfo{author}{\bibfnamefont{Y.}~\bibnamefont{Sekiguchi}},
  \bibinfo{author}{\bibfnamefont{K.}~\bibnamefont{Kiuchi}},
  \bibinfo{author}{\bibfnamefont{K.}~\bibnamefont{Kyutoku}}, \bibnamefont{and}
  \bibinfo{author}{\bibfnamefont{M.}~\bibnamefont{Shibata}},
  \bibinfo{journal}{Phys. Rev. D} \textbf{\bibinfo{volume}{91}},
  \bibinfo{pages}{064059} (\bibinfo{year}{2015}). 

\bibitem[{\citenamefont{{Lattimer} and {Swesty}}(1991)}]{lattimer91}
\bibinfo{author}{\bibfnamefont{J.~M.} \bibnamefont{{Lattimer}}}
  \bibnamefont{and} \bibinfo{author}{\bibfnamefont{F.~D.}
  \bibnamefont{{Swesty}}}, \bibinfo{journal}{Nuclear Physics A}
  \textbf{\bibinfo{volume}{535}}, \bibinfo{pages}{331} (\bibinfo{year}{1991}).

\bibitem[{\citenamefont{{Shen} et~al.}(1998{\natexlab{a}})\citenamefont{{Shen},
  {Toki}, {Oyamatsu}, and {Sumiyoshi}}}]{shen98a}
\bibinfo{author}{\bibfnamefont{H.}~\bibnamefont{{Shen}}},
  \bibinfo{author}{\bibfnamefont{H.}~\bibnamefont{{Toki}}},
  \bibinfo{author}{\bibfnamefont{K.}~\bibnamefont{{Oyamatsu}}},
  \bibnamefont{and}
  \bibinfo{author}{\bibfnamefont{K.}~\bibnamefont{{Sumiyoshi}}},
  \bibinfo{journal}{Nuclear Physics A} \textbf{\bibinfo{volume}{637}},
  \bibinfo{pages}{435} (\bibinfo{year}{1998}{\natexlab{a}}).

\bibitem[{\citenamefont{{Shen} et~al.}(1998{\natexlab{b}})\citenamefont{{Shen},
  {Toki}, {Oyamatsu}, and {Sumiyoshi}}}]{shen98b}
\bibinfo{author}{\bibfnamefont{H.}~\bibnamefont{{Shen}}},
  \bibinfo{author}{\bibfnamefont{H.}~\bibnamefont{{Toki}}},
  \bibinfo{author}{\bibfnamefont{K.}~\bibnamefont{{Oyamatsu}}},
  \bibnamefont{and}
  \bibinfo{author}{\bibfnamefont{K.}~\bibnamefont{{Sumiyoshi}}},
  \bibinfo{journal}{Progress of Theoretical Physics}
  \textbf{\bibinfo{volume}{100}}, \bibinfo{pages}{1013}
  (\bibinfo{year}{1998}{\natexlab{b}}).

\bibitem[{\citenamefont{Shen et~al.}(2011)\citenamefont{Shen, Toki, Oyamatsu,
  and Sumiyoshi}}]{shen11}
\bibinfo{author}{\bibfnamefont{H.}~\bibnamefont{Shen}},
  \bibinfo{author}{\bibfnamefont{H.}~\bibnamefont{Toki}},
  \bibinfo{author}{\bibfnamefont{K.}~\bibnamefont{Oyamatsu}}, \bibnamefont{and}
  \bibinfo{author}{\bibfnamefont{K.}~\bibnamefont{Sumiyoshi}},
  \bibinfo{journal}{Astrophys.J.Suppl.} \textbf{\bibinfo{volume}{197}},
  \bibinfo{pages}{20} (\bibinfo{year}{2011}).

\bibitem[{\citenamefont{{Burrows} and {Lattimer}}(1984)}]{burrows84}
\bibinfo{author}{\bibfnamefont{A.}~\bibnamefont{{Burrows}}} \bibnamefont{and}
  \bibinfo{author}{\bibfnamefont{J.~M.} \bibnamefont{{Lattimer}}},
  \bibinfo{journal}{\apj} \textbf{\bibinfo{volume}{285}}, \bibinfo{pages}{294}
  (\bibinfo{year}{1984}).

\bibitem[{\citenamefont{{Hempel} and {Schaffner-Bielich}}(2010)}]{hempel10}
\bibinfo{author}{\bibfnamefont{M.}~\bibnamefont{{Hempel}}} \bibnamefont{and}
  \bibinfo{author}{\bibfnamefont{J.}~\bibnamefont{{Schaffner-Bielich}}},
  \bibinfo{journal}{Nuclear Physics A} \textbf{\bibinfo{volume}{837}},
  \bibinfo{pages}{210} (\bibinfo{year}{2010}).

\bibitem[{\citenamefont{{Furusawa} et~al.}(2011)\citenamefont{{Furusawa},
  {Yamada}, {Sumiyoshi}, and {Suzuki}}}]{furusawa11}
\bibinfo{author}{\bibfnamefont{S.}~\bibnamefont{{Furusawa}}},
  \bibinfo{author}{\bibfnamefont{S.}~\bibnamefont{{Yamada}}},
  \bibinfo{author}{\bibfnamefont{K.}~\bibnamefont{{Sumiyoshi}}},
  \bibnamefont{and} \bibinfo{author}{\bibfnamefont{H.}~\bibnamefont{{Suzuki}}},
  \bibinfo{journal}{\apj} \textbf{\bibinfo{volume}{738}}, \bibinfo{eid}{178}
  (\bibinfo{year}{2011}).

\bibitem[{\citenamefont{{Botvina} and {Mishustin}}(2004)}]{botvina04}
\bibinfo{author}{\bibfnamefont{A.~S.} \bibnamefont{{Botvina}}}
  \bibnamefont{and} \bibinfo{author}{\bibfnamefont{I.~N.}
  \bibnamefont{{Mishustin}}}, \bibinfo{journal}{Physics Letters B}
  \textbf{\bibinfo{volume}{584}}, \bibinfo{pages}{233} (\bibinfo{year}{2004}) .

\bibitem[{\citenamefont{{Botvina} and {Mishustin}}(2010)}]{botvina10}
\bibinfo{author}{\bibfnamefont{A.~S.} \bibnamefont{{Botvina}}}
  \bibnamefont{and} \bibinfo{author}{\bibfnamefont{I.~N.}
  \bibnamefont{{Mishustin}}}, \bibinfo{journal}{Nuclear Physics A}
  \textbf{\bibinfo{volume}{843}}, \bibinfo{pages}{98} (\bibinfo{year}{2010}). 

\bibitem[{\citenamefont{{Buyukcizmeci}
  et~al.}(2014)\citenamefont{{Buyukcizmeci}, {Botvina}, and
  {Mishustin}}}]{buyukcizmeci14}
\bibinfo{author}{\bibfnamefont{N.}~\bibnamefont{{Buyukcizmeci}}},
  \bibinfo{author}{\bibfnamefont{A.~S.} \bibnamefont{{Botvina}}},
  \bibnamefont{and} \bibinfo{author}{\bibfnamefont{I.~N.}
  \bibnamefont{{Mishustin}}}, \bibinfo{journal}{\apj}
  \textbf{\bibinfo{volume}{789}}, \bibinfo{eid}{33} (\bibinfo{year}{2014}). 

\bibitem[{\citenamefont{{Bondorf} et~al.}(1995)\citenamefont{{Bondorf},
  {Botvina}, {Iljinov}, {Mishustin}, and {Sneppen}}}]{bondorf95}
\bibinfo{author}{\bibfnamefont{J.~P.} \bibnamefont{{Bondorf}}},
  \bibinfo{author}{\bibfnamefont{A.~S.} \bibnamefont{{Botvina}}},
  \bibinfo{author}{\bibfnamefont{A.~S.} \bibnamefont{{Iljinov}}},
  \bibinfo{author}{\bibfnamefont{I.~N.} \bibnamefont{{Mishustin}}},
  \bibnamefont{and}
  \bibinfo{author}{\bibfnamefont{K.}~\bibnamefont{{Sneppen}}},
  \bibinfo{journal}{\physrep} \textbf{\bibinfo{volume}{257}},
  \bibinfo{pages}{133} (\bibinfo{year}{1995}).

\bibitem[{\citenamefont{{Furusawa}
  et~al.}(2013{\natexlab{a}})\citenamefont{{Furusawa}, {Sumiyoshi}, {Yamada},
  and {Suzuki}}}]{furusawa13a}
\bibinfo{author}{\bibfnamefont{S.}~\bibnamefont{{Furusawa}}},
  \bibinfo{author}{\bibfnamefont{K.}~\bibnamefont{{Sumiyoshi}}},
  \bibinfo{author}{\bibfnamefont{S.}~\bibnamefont{{Yamada}}}, \bibnamefont{and}
  \bibinfo{author}{\bibfnamefont{H.}~\bibnamefont{{Suzuki}}},
  \bibinfo{journal}{\apj} \textbf{\bibinfo{volume}{772}}, \bibinfo{eid}{95}
  (\bibinfo{year}{2013}{\natexlab{a}}).

\bibitem[{\citenamefont{Furusawa et~al.}(2017)\citenamefont{Furusawa,
  Sumiyoshi, Yamada, and Suzuki}}]{furusawa17a}
\bibinfo{author}{\bibfnamefont{S.}~\bibnamefont{Furusawa}},
  \bibinfo{author}{\bibfnamefont{K.}~\bibnamefont{Sumiyoshi}},
  \bibinfo{author}{\bibfnamefont{S.}~\bibnamefont{Yamada}}, \bibnamefont{and}
  \bibinfo{author}{\bibfnamefont{H.}~\bibnamefont{Suzuki}},
  \bibinfo{journal}{Nuclear Physics A} \textbf{\bibinfo{volume}{957}},
  \bibinfo{pages}{188 } (\bibinfo{year}{2017}), ISSN \bibinfo{issn}{0375-9474}.

\bibitem[{\citenamefont{{Buyukcizmeci}
  et~al.}(2013)\citenamefont{{Buyukcizmeci}, {Botvina}, {Mishustin}, {Ogul},
  {Hempel}, {Schaffner-Bielich}, {Thielemann}, {Furusawa}, {Sumiyoshi},
  {Yamada} et~al.}}]{buyukcizmeci13}
\bibinfo{author}{\bibfnamefont{N.}~\bibnamefont{{Buyukcizmeci}}},
  \bibinfo{author}{\bibfnamefont{A.~S.} \bibnamefont{{Botvina}}},
  \bibinfo{author}{\bibfnamefont{I.~N.} \bibnamefont{{Mishustin}}},
  \bibinfo{author}{\bibfnamefont{R.}~\bibnamefont{{Ogul}}},
  \bibinfo{author}{\bibfnamefont{M.}~\bibnamefont{{Hempel}}},
  \bibinfo{author}{\bibfnamefont{J.}~\bibnamefont{{Schaffner-Bielich}}},
  \bibinfo{author}{\bibfnamefont{F.-K.} \bibnamefont{{Thielemann}}},
  \bibinfo{author}{\bibfnamefont{S.}~\bibnamefont{{Furusawa}}},
  \bibinfo{author}{\bibfnamefont{K.}~\bibnamefont{{Sumiyoshi}}},
  \bibinfo{author}{\bibfnamefont{S.}~\bibnamefont{{Yamada}}},
  \bibnamefont{et~al.}, \bibinfo{journal}{Nuclear Physics A}
  \textbf{\bibinfo{volume}{907}}, \bibinfo{pages}{13} (\bibinfo{year}{2013}). 

\bibitem[{\citenamefont{{Shen} et~al.}(2011)\citenamefont{{Shen}, {Horowitz},
  and {Teige}}}]{sheng11}
\bibinfo{author}{\bibfnamefont{G.}~\bibnamefont{{Shen}}},
  \bibinfo{author}{\bibfnamefont{C.~J.} \bibnamefont{{Horowitz}}},
  \bibnamefont{and} \bibinfo{author}{\bibfnamefont{S.}~\bibnamefont{{Teige}}},
  \bibinfo{journal}{\prc} \textbf{\bibinfo{volume}{83}}, \bibinfo{eid}{035802}
  (\bibinfo{year}{2011}).

\bibitem[{\citenamefont{{Typel}}(2015)}]{typelpc}
\bibinfo{author}{\bibfnamefont{S.}~\bibnamefont{{Typel}}},
  \bibinfo{journal}{private communication}  (\bibinfo{year}{2015}).

\bibitem[{\citenamefont{Papakonstantinou
  et~al.}(2013)\citenamefont{Papakonstantinou, Margueron, Gulminelli, and
  Raduta}}]{papakonstantinou13}
\bibinfo{author}{\bibfnamefont{P.}~\bibnamefont{Papakonstantinou}},
  \bibinfo{author}{\bibfnamefont{J.}~\bibnamefont{Margueron}},
  \bibinfo{author}{\bibfnamefont{F.}~\bibnamefont{Gulminelli}},
  \bibnamefont{and} \bibinfo{author}{\bibfnamefont{A.~R.}
  \bibnamefont{Raduta}}, \bibinfo{journal}{Phys. Rev. C}
  \textbf{\bibinfo{volume}{88}}, \bibinfo{pages}{045805}
  (\bibinfo{year}{2013}).

\bibitem[{\citenamefont{{Aymard} et~al.}(2014)\citenamefont{{Aymard},
  {Gulminelli}, and {Margueron}}}]{aymard14}
\bibinfo{author}{\bibfnamefont{F.}~\bibnamefont{{Aymard}}},
  \bibinfo{author}{\bibfnamefont{F.}~\bibnamefont{{Gulminelli}}},
  \bibnamefont{and}
  \bibinfo{author}{\bibfnamefont{J.}~\bibnamefont{{Margueron}}},
  \bibinfo{journal}{\prc} \textbf{\bibinfo{volume}{89}}, \bibinfo{eid}{065807}
  (\bibinfo{year}{2014}).

\bibitem[{\citenamefont{Gulminelli and Raduta}(2015)}]{gulminelli15}
\bibinfo{author}{\bibfnamefont{F.}~\bibnamefont{Gulminelli}} \bibnamefont{and}
  \bibinfo{author}{\bibfnamefont{A.~R.} \bibnamefont{Raduta}},
  \bibinfo{journal}{Phys. Rev. C} \textbf{\bibinfo{volume}{92}},
  \bibinfo{pages}{055803} (\bibinfo{year}{2015}). 

\bibitem[{\citenamefont{{Furusawa}
  et~al.}(2013{\natexlab{b}})\citenamefont{{Furusawa}, {Nagakura}, {Sumiyoshi},
  and {Yamada}}}]{furusawa13b}
\bibinfo{author}{\bibfnamefont{S.}~\bibnamefont{{Furusawa}}},
  \bibinfo{author}{\bibfnamefont{H.}~\bibnamefont{{Nagakura}}},
  \bibinfo{author}{\bibfnamefont{K.}~\bibnamefont{{Sumiyoshi}}},
  \bibnamefont{and} \bibinfo{author}{\bibfnamefont{S.}~\bibnamefont{{Yamada}}},
  \bibinfo{journal}{\apj} \textbf{\bibinfo{volume}{774}}, \bibinfo{eid}{78}
  (\bibinfo{year}{2013}{\natexlab{b}}).

\bibitem[{\citenamefont{{Fischer} et~al.}(2016)\citenamefont{{Fischer},
  {Mart{\'{\i}}nez-Pinedo}, {Hempel}, {Huther}, {R{\"o}pke}, {Typel}, and
  {Lohs}}}]{fischer16}
\bibinfo{author}{\bibfnamefont{T.}~\bibnamefont{{Fischer}}},
  \bibinfo{author}{\bibfnamefont{G.}~\bibnamefont{{Mart{\'{\i}}nez-Pinedo}}},
  \bibinfo{author}{\bibfnamefont{M.}~\bibnamefont{{Hempel}}},
  \bibinfo{author}{\bibfnamefont{L.}~\bibnamefont{{Huther}}},
  \bibinfo{author}{\bibfnamefont{G.}~\bibnamefont{{R{\"o}pke}}},
  \bibinfo{author}{\bibfnamefont{S.}~\bibnamefont{{Typel}}}, \bibnamefont{and}
  \bibinfo{author}{\bibfnamefont{A.}~\bibnamefont{{Lohs}}}, 
  \bibinfo{journal}{European Physical Journal Web of Conferences},
 \textbf{\bibinfo{volume}{109}}, \bibinfo{pages}{06002}  (\bibinfo{year}{2016}).

\bibitem[{\citenamefont{Hempel et~al.}(2011)\citenamefont{Hempel,
  Schaffner-Bielich, Typel, and R\"opke}}]{hempel11}
\bibinfo{author}{\bibfnamefont{M.}~\bibnamefont{Hempel}},
  \bibinfo{author}{\bibfnamefont{J.}~\bibnamefont{Schaffner-Bielich}},
  \bibinfo{author}{\bibfnamefont{S.}~\bibnamefont{Typel}}, \bibnamefont{and}
  \bibinfo{author}{\bibfnamefont{G.}~\bibnamefont{R\"opke}},
  \bibinfo{journal}{Phys. Rev. C} \textbf{\bibinfo{volume}{84}},
  \bibinfo{pages}{055804} (\bibinfo{year}{2011}),

\bibitem[{\citenamefont{{Oertel} et~al.}(2016)\citenamefont{{Oertel}, {Hempel},
  {Kl{\"a}hn}, and {Typel}}}]{oertel16}
\bibinfo{author}{\bibfnamefont{M.}~\bibnamefont{{Oertel}}},
  \bibinfo{author}{\bibfnamefont{M.}~\bibnamefont{{Hempel}}},
  \bibinfo{author}{\bibfnamefont{T.}~\bibnamefont{{Kl{\"a}hn}}},
  \bibnamefont{and} \bibinfo{author}{\bibfnamefont{S.}~\bibnamefont{{Typel}}},
  \bibinfo{journal}{ArXiv e-prints}  (\bibinfo{year}{2016})   \eprint{1610.03361}.

\bibitem[{\citenamefont{Oyamatsu and Iida}(2003)}]{oyamatsu03}
\bibinfo{author}{\bibfnamefont{K.}~\bibnamefont{Oyamatsu}} \bibnamefont{and}
  \bibinfo{author}{\bibfnamefont{K.}~\bibnamefont{Iida}},
  \bibinfo{journal}{Prog. Theor. Phys.} \textbf{\bibinfo{volume}{109}},
  \bibinfo{pages}{631} (\bibinfo{year}{2003}).

\bibitem[{\citenamefont{{Oyamatsu} and {Iida}}(2007)}]{oyamatsu07}
\bibinfo{author}{\bibfnamefont{K.}~\bibnamefont{{Oyamatsu}}} \bibnamefont{and}
  \bibinfo{author}{\bibfnamefont{K.}~\bibnamefont{{Iida}}},
  \bibinfo{journal}{\prc} \textbf{\bibinfo{volume}{75}}, \bibinfo{eid}{015801}
  (\bibinfo{year}{2007}).

\bibitem[{\citenamefont{Tsang et~al.}(2012)\citenamefont{Tsang, Stone, Camera,
  Danielewicz, Gandolfi, Hebeler, Horowitz, Lee, Lynch, Kohley
  et~al.}}]{tsang12}
\bibinfo{author}{\bibfnamefont{M.~B.} \bibnamefont{Tsang}},
  \bibinfo{author}{\bibfnamefont{J.~R.} \bibnamefont{Stone}},
  \bibinfo{author}{\bibfnamefont{F.}~\bibnamefont{Camera}},
  \bibinfo{author}{\bibfnamefont{P.}~\bibnamefont{Danielewicz}},
  \bibinfo{author}{\bibfnamefont{S.}~\bibnamefont{Gandolfi}},
  \bibinfo{author}{\bibfnamefont{K.}~\bibnamefont{Hebeler}},
  \bibinfo{author}{\bibfnamefont{C.~J.} \bibnamefont{Horowitz}},
  \bibinfo{author}{\bibfnamefont{J.}~\bibnamefont{Lee}},
  \bibinfo{author}{\bibfnamefont{W.~G.} \bibnamefont{Lynch}},
  \bibinfo{author}{\bibfnamefont{Z.}~\bibnamefont{Kohley}},
  \bibnamefont{et~al.}, \bibinfo{journal}{Phys. Rev. C}
  \textbf{\bibinfo{volume}{86}}, \bibinfo{pages}{015803}
  (\bibinfo{year}{2012}). 

\bibitem[{\citenamefont{Newton et~al.}(2013)\citenamefont{Newton, Gearheart,
  Wen, and Li}}]{newton13}
\bibinfo{author}{\bibfnamefont{W.~G.} \bibnamefont{Newton}},
  \bibinfo{author}{\bibfnamefont{M.}~\bibnamefont{Gearheart}},
  \bibinfo{author}{\bibfnamefont{D.-H.} \bibnamefont{Wen}}, \bibnamefont{and}
  \bibinfo{author}{\bibfnamefont{B.-A.} \bibnamefont{Li}},
  \bibinfo{journal}{Journal of Physics: Conference Series}
  \textbf{\bibinfo{volume}{420}}, \bibinfo{pages}{012145}
  (\bibinfo{year}{2013}). 

\bibitem[{\citenamefont{Lattimer and Steiner}(2014)}]{lattimer14a}
\bibinfo{author}{\bibfnamefont{J.~M.} \bibnamefont{Lattimer}} \bibnamefont{and}
  \bibinfo{author}{\bibfnamefont{A.~W.} \bibnamefont{Steiner}},
  \bibinfo{journal}{The European Physical Journal A}
  \textbf{\bibinfo{volume}{50}}, \bibinfo{pages}{1} (\bibinfo{year}{2014}),
  ISSN \bibinfo{issn}{1434-601X}. 

\bibitem[{\citenamefont{Hebeler and Schwenk}(2014)}]{hebeler14}
\bibinfo{author}{\bibfnamefont{K.}~\bibnamefont{Hebeler}} \bibnamefont{and}
  \bibinfo{author}{\bibfnamefont{A.}~\bibnamefont{Schwenk}},
  \bibinfo{journal}{The European Physical Journal A}
  \textbf{\bibinfo{volume}{50}}, \bibinfo{pages}{1} (\bibinfo{year}{2014}),
  ISSN \bibinfo{issn}{1434-601X}. 

\bibitem[{\citenamefont{Danielewicz and Lee}(2014)}]{danielewicz14}
\bibinfo{author}{\bibfnamefont{P.}~\bibnamefont{Danielewicz}} \bibnamefont{and}
  \bibinfo{author}{\bibfnamefont{J.}~\bibnamefont{Lee}},
  \bibinfo{journal}{Nuclear Physics A} \textbf{\bibinfo{volume}{922}},
  \bibinfo{pages}{1 } (\bibinfo{year}{2014}), ISSN \bibinfo{issn}{0375-9474}. 

\bibitem[{\citenamefont{Sotani et~al.}(2015)\citenamefont{Sotani, Iida, and
  Oyamatsu}}]{sotani15}
\bibinfo{author}{\bibfnamefont{H.}~\bibnamefont{Sotani}},
  \bibinfo{author}{\bibfnamefont{K.}~\bibnamefont{Iida}}, \bibnamefont{and}
  \bibinfo{author}{\bibfnamefont{K.}~\bibnamefont{Oyamatsu}},
  \bibinfo{journal}{Phys. Rev. C} \textbf{\bibinfo{volume}{91}},
  \bibinfo{pages}{015805} (\bibinfo{year}{2015}). 

\bibitem[{\citenamefont{{Shlomo} et~al.}(2006)\citenamefont{{Shlomo},
  {Kolomietz}, and {Col{\`o}}}}]{shlomo06}
\bibinfo{author}{\bibfnamefont{S.}~\bibnamefont{{Shlomo}}},
  \bibinfo{author}{\bibfnamefont{V.~M.} \bibnamefont{{Kolomietz}}},
  \bibnamefont{and}
  \bibinfo{author}{\bibfnamefont{G.}~\bibnamefont{{Col{\`o}}}},
  \bibinfo{journal}{European Physical Journal A} \textbf{\bibinfo{volume}{30}},
  \bibinfo{pages}{23} (\bibinfo{year}{2006}).

\bibitem[{\citenamefont{{Lattimer} and {Lim}}(2013)}]{lattimer13}
\bibinfo{author}{\bibfnamefont{J.~M.} \bibnamefont{{Lattimer}}}
  \bibnamefont{and} \bibinfo{author}{\bibfnamefont{Y.}~\bibnamefont{{Lim}}},
  \bibinfo{journal}{\apj} \textbf{\bibinfo{volume}{771}}, \bibinfo{eid}{51}
  (\bibinfo{year}{2013}).

\bibitem[{\citenamefont{{Watanabe} et~al.}(2005)\citenamefont{{Watanabe},
  {Maruyama}, {Sato}, {Yasuoka}, and {Ebisuzaki}}}]{watanabe05}
\bibinfo{author}{\bibfnamefont{G.}~\bibnamefont{{Watanabe}}},
  \bibinfo{author}{\bibfnamefont{T.}~\bibnamefont{{Maruyama}}},
  \bibinfo{author}{\bibfnamefont{K.}~\bibnamefont{{Sato}}},
  \bibinfo{author}{\bibfnamefont{K.}~\bibnamefont{{Yasuoka}}},
  \bibnamefont{and}
  \bibinfo{author}{\bibfnamefont{T.}~\bibnamefont{{Ebisuzaki}}},
  \bibinfo{journal}{Physical Review Letters} \textbf{\bibinfo{volume}{94}},
  \bibinfo{eid}{031101} (\bibinfo{year}{2005}).

\bibitem[{\citenamefont{{Newton} and {Stone}}(2009)}]{newton09}
\bibinfo{author}{\bibfnamefont{W.~G.} \bibnamefont{{Newton}}} \bibnamefont{and}
  \bibinfo{author}{\bibfnamefont{J.~R.} \bibnamefont{{Stone}}},
  \bibinfo{journal}{\prc} \textbf{\bibinfo{volume}{79}}, \bibinfo{eid}{055801}
  (\bibinfo{year}{2009}).

\bibitem[{\citenamefont{{Okamoto} et~al.}(2012)\citenamefont{{Okamoto},
  {Maruyama}, {Yabana}, and {Tatsumi}}}]{okamoto12}
\bibinfo{author}{\bibfnamefont{M.}~\bibnamefont{{Okamoto}}},
  \bibinfo{author}{\bibfnamefont{T.}~\bibnamefont{{Maruyama}}},
  \bibinfo{author}{\bibfnamefont{K.}~\bibnamefont{{Yabana}}}, \bibnamefont{and}
  \bibinfo{author}{\bibfnamefont{T.}~\bibnamefont{{Tatsumi}}},
  \bibinfo{journal}{Physics Letters B} \textbf{\bibinfo{volume}{713}},
  \bibinfo{pages}{284} (\bibinfo{year}{2012}).

\bibitem[{\citenamefont{Schneider et~al.}(2013)\citenamefont{Schneider,
  Horowitz, Hughto, and Berry}}]{schneider13}
\bibinfo{author}{\bibfnamefont{A.~S.} \bibnamefont{Schneider}},
  \bibinfo{author}{\bibfnamefont{C.~J.} \bibnamefont{Horowitz}},
  \bibinfo{author}{\bibfnamefont{J.}~\bibnamefont{Hughto}}, \bibnamefont{and}
  \bibinfo{author}{\bibfnamefont{D.~K.} \bibnamefont{Berry}},
  \bibinfo{journal}{Phys. Rev. C} \textbf{\bibinfo{volume}{88}},
  \bibinfo{pages}{065807} (\bibinfo{year}{2013}). 

\bibitem[{\citenamefont{{Horowitz} et~al.}(2016)\citenamefont{{Horowitz},
  {Berry}, {Caplan}, {Fischer}, {Lin}, {Newton}, {O'Connor}, and
  {Roberts}}}]{horowitz16}
\bibinfo{author}{\bibfnamefont{C.~J.} \bibnamefont{{Horowitz}}},
  \bibinfo{author}{\bibfnamefont{D.~K.} \bibnamefont{{Berry}}},
  \bibinfo{author}{\bibfnamefont{M.~E.} \bibnamefont{{Caplan}}},
  \bibinfo{author}{\bibfnamefont{T.}~\bibnamefont{{Fischer}}},
  \bibinfo{author}{\bibfnamefont{Z.}~\bibnamefont{{Lin}}},
  \bibinfo{author}{\bibfnamefont{W.~G.} \bibnamefont{{Newton}}},
  \bibinfo{author}{\bibfnamefont{E.}~\bibnamefont{{O'Connor}}},
  \bibnamefont{and} \bibinfo{author}{\bibfnamefont{L.~F.}
  \bibnamefont{{Roberts}}}, \bibinfo{journal}{ArXiv e-prints}
  (\bibinfo{year}{2016}).

\bibitem[{\citenamefont{Maruyama et~al.}(2005)\citenamefont{Maruyama, Tatsumi,
  Voskresensky, Tanigawa, and Chiba}}]{maruyama05}
\bibinfo{author}{\bibfnamefont{T.}~\bibnamefont{Maruyama}},
  \bibinfo{author}{\bibfnamefont{T.}~\bibnamefont{Tatsumi}},
  \bibinfo{author}{\bibfnamefont{D.~N.} \bibnamefont{Voskresensky}},
  \bibinfo{author}{\bibfnamefont{T.}~\bibnamefont{Tanigawa}}, \bibnamefont{and}
  \bibinfo{author}{\bibfnamefont{S.}~\bibnamefont{Chiba}},
  \bibinfo{journal}{Phys. Rev. C} \textbf{\bibinfo{volume}{72}},
  \bibinfo{pages}{015802} (\bibinfo{year}{2005}). 

\bibitem[{\citenamefont{Ebel et~al.}(2015)\citenamefont{Ebel, Mishustin, and
  Greiner}}]{ebel15}
\bibinfo{author}{\bibfnamefont{C.}~\bibnamefont{Ebel}},
  \bibinfo{author}{\bibfnamefont{I.}~\bibnamefont{Mishustin}},
  \bibnamefont{and} \bibinfo{author}{\bibfnamefont{W.}~\bibnamefont{Greiner}},
  \bibinfo{journal}{Journal of Physics G: Nuclear and Particle Physics}
  \textbf{\bibinfo{volume}{42}}, \bibinfo{pages}{105201}
  (\bibinfo{year}{2015}). 

\bibitem[{\citenamefont{{Agrawal} et~al.}(2014)\citenamefont{{Agrawal},
  {Bandyopadhyay}, {De}, and {Samaddar}}}]{agrawal14b}
\bibinfo{author}{\bibfnamefont{B.~K.} \bibnamefont{{Agrawal}}},
  \bibinfo{author}{\bibfnamefont{D.}~\bibnamefont{{Bandyopadhyay}}},
  \bibinfo{author}{\bibfnamefont{J.~N.} \bibnamefont{{De}}}, \bibnamefont{and}
  \bibinfo{author}{\bibfnamefont{S.~K.} \bibnamefont{{Samaddar}}},
  \bibinfo{journal}{\prc} \textbf{\bibinfo{volume}{89}}, \bibinfo{eid}{044320}
  (\bibinfo{year}{2014}).

\bibitem[{\citenamefont{{Sullivan} et~al.}(2016)\citenamefont{{Sullivan},
  {O'Connor}, {Zegers}, {Grubb}, and {Austin}}}]{sullivan16}
\bibinfo{author}{\bibfnamefont{C.}~\bibnamefont{{Sullivan}}},
  \bibinfo{author}{\bibfnamefont{E.}~\bibnamefont{{O'Connor}}},
  \bibinfo{author}{\bibfnamefont{R.~G.~T.} \bibnamefont{{Zegers}}},
  \bibinfo{author}{\bibfnamefont{T.}~\bibnamefont{{Grubb}}}, \bibnamefont{and}
  \bibinfo{author}{\bibfnamefont{S.~M.} \bibnamefont{{Austin}}},
  \bibinfo{journal}{\apj} \textbf{\bibinfo{volume}{816}}, \bibinfo{eid}{44}
  (\bibinfo{year}{2016}).

\bibitem[{\citenamefont{Sekiguchi et~al.}(2016)\citenamefont{Sekiguchi, Kiuchi,
  Kyutoku, Shibata, and Taniguchi}}]{sekiguchi16}
\bibinfo{author}{\bibfnamefont{Y.}~\bibnamefont{Sekiguchi}},
  \bibinfo{author}{\bibfnamefont{K.}~\bibnamefont{Kiuchi}},
  \bibinfo{author}{\bibfnamefont{K.}~\bibnamefont{Kyutoku}},
  \bibinfo{author}{\bibfnamefont{M.}~\bibnamefont{Shibata}}, \bibnamefont{and}
  \bibinfo{author}{\bibfnamefont{K.}~\bibnamefont{Taniguchi}},
  \bibinfo{journal}{Phys. Rev. D} \textbf{\bibinfo{volume}{93}},
  \bibinfo{pages}{124046} (\bibinfo{year}{2016}). 

\bibitem[{\citenamefont{Sumiyoshi and R\"opke}(2008)}]{sumiyoshi08}
\bibinfo{author}{\bibfnamefont{K.}~\bibnamefont{Sumiyoshi}} \bibnamefont{and}
  \bibinfo{author}{\bibfnamefont{G.}~\bibnamefont{R\"opke}},
  \bibinfo{journal}{Phys. Rev. C} \textbf{\bibinfo{volume}{77}},
  \bibinfo{pages}{055804} (\bibinfo{year}{2008}). 

\end{thebibliography}

\newpage

\begin{table}[t]
\begin{tabular}{|c||c|c|c|c|c|}
\hline
   parameter set  &  $n_{0}$ (fm$^{-3}$)  & $\omega_0$ (MeV) & $K_0$ (MeV) & $S_0$ (MeV)  & $L$ (MeV) \\
 \hline
 \hline
 B  &  0.15969 & -16.184  &  230   & 33.550 & 73.214 \\
 \hline
 E & 0.15979 &  -16.145  &  230  & 31.002 & 42.498  \\
\hline
\end{tabular}
\caption{\label{tab1_bulk}%
Bulk properties of nuclear matter \cite{oyamatsu03,oyamatsu07}. }  
\end{table}

\end{document}